\documentclass{article}

\usepackage{PRIMEarxiv}

\usepackage[utf8]{inputenc} 
\usepackage[T1]{fontenc}    
\usepackage{url}            
\usepackage{booktabs}       
\usepackage{amsfonts}       
\usepackage{nicefrac}       
\usepackage{microtype}      
\usepackage{lipsum}
\usepackage{fancyhdr}       
\usepackage{graphicx}       

\newcommand{\methodsref}[1]{\hyperref[#1]{Methods}}

\usepackage[title]{appendix}
\usepackage{amsmath}
\usepackage{amsfonts}
\usepackage{amssymb}
\usepackage{graphicx}
\usepackage{siunitx}
\usepackage{xcolor}
\usepackage{booktabs}
\usepackage{caption}
\usepackage{float}
\usepackage{color,soul}
\usepackage{hyperref}
\usepackage{makecell}
\usepackage{xurl}
\usepackage{algorithm} 
\usepackage{algpseudocode}

\hypersetup{
    colorlinks=true,
    linkcolor=blue,
    filecolor=blue,      
    urlcolor=blue,
    pdftitle={Overleaf Example},
    pdfpagemode=FullScreen,
    }

\usepackage{bm}

\usepackage{comment}

\title{Learning Turbulent Flows with Generative Models: Super-resolution, Forecasting, and \\Sparse Flow Reconstruction}

\author{
  Vivek Oommen$^{1}$, Siavash Khodakarami$^{2}$, Aniruddha Bora$^2$, Zhicheng Wang$^{2}$, George Em Karniadakis$^2$\thanks{Corresponding author: george\_karniadakis@brown.edu}  \\ \\
  1-School of Engineering \\
  2-Division of Applied Mathematics \\
  Brown University \\
  Providence, RI 02912\\
}

\begin{document}
\maketitle

\begin{abstract}
Neural operators are promising surrogates for dynamical systems but when trained with standard $L^2$ losses they tend to oversmooth fine-scale turbulent structures. 
Here, we show that combining operator learning with generative modeling overcomes this limitation.
We consider three practical turbulent-flow challenges where conventional neural operators fail: spatio-temporal super-resolution, forecasting, and sparse flow reconstruction.
For Schlieren jet super-resolution, an adversarially trained neural operator (adv-NO) reduces the energy-spectrum error by $15\times$ while preserving sharp gradients at neural operator-like inference cost. 
For 3D homogeneous isotropic turbulence, adv-NO trained on only 160 timesteps from a single trajectory forecasts accurately for five eddy-turnover times and offers $114\times$ wall-clock speed-up at inference than the baseline diffusion-based forecasters, enabling near-real-time rollouts. 
For reconstructing cylinder wake flows from highly sparse Particle Tracking Velocimetry-like inputs, a conditional generative model infers full 3D velocity and pressure fields with correct phase alignment and statistics. 
These advances enable accurate reconstruction and forecasting at low compute cost, bringing near-real-time analysis and control within reach in experimental and computational fluid mechanics. \newline
See our project page: \url{https://vivekoommen.github.io/Gen4Turb/}
\end{abstract}


\section*{Introduction}
\label{sec: intro}
In this work, we explore how generative modeling and operator learning can be applied synergistically to address three central challenges in turbulence: spatio-temporal super-resolution, forecasting with limited data, and reconstruction from sparse observations.
Our approach combines adversarial training with operator learning for super-resolution and forecasting.
We further leverage generative models to reconstruct turbulent flows from extremely sparse Particle Tracking Velocimetry (PTV)-like velocity measurements, achieving phase alignment and matching flow statistics even when trained with as few as 150 snapshots.
To place these contributions in context, we first recall how turbulence modeling and simulation have evolved over the past decades.

{
In the early 1970s, Howard W. Emmons argued that direct numerical simulation (DNS) of turbulence would never be computable with standard Fourier methods, because the quadratic nonlinearity in Navier-Stokes would inevitably alias high‐wavenumber energy back into the resolved band \cite{emmons1970critique}. 
That pessimism was overturned when Patterson \& Orszag introduced the simple 3/2‐rule: zero‐pad the Fourier modes to 1.5 N in each direction, multiply in physical space, then truncate back to N, thereby removing all aliasing and enabling the first practical DNS of homogeneous isotropic turbulence on a $(2\pi)^3$ grid \cite{orszag1972numerical,orszag1979spectral} in 1972.  
These ideas were later extended to complex-geometry flows using the spectral element method, see e.g.,  \cite{karniadakis1993nodes} in 1993.
However, simulating turbulent flows remains computationally expensive due to the need for fine temporal discretizations in DNS as well as relatively cheaper large-eddy simulations (LES). 
These small time steps are typically required to ensure numerical stability and accurately resolve the nonlinear dynamics of turbulence. 
In contrast, neural operators (NOs) \cite{lu2021learning,li2020fourier} offer a promising alternative for modeling turbulent flows by effectively learning non-linear mappings across larger time intervals, potentially bypassing restrictive time-stepping constraints.
}

%


{
Neural networks, and more recently neural operators, have gained traction in scientific computing due to their ability to approximate complex input-output functional mappings with fast inference once trained offline. 
They have been applied across a range of domains - from fluid dynamics and weather forecasting to materials science and quantum chemistry. 
Theoretical results like the universal operator approximation theorem by Chen and Chen \cite{chen1995universal} have further accelerated interest by guaranteeing that neural operators can learn mappings between infinite-dimensional function spaces. 
Several architectures have been proposed, including the Deep Operator Network (DeepONet) \cite{lu2021learning}, which uses branch and trunk networks to learn functionals; the Fourier and Laplace Neural Operators (FNO \& LNO) \cite{li2020fourier,cao2024laplace}, which use global convolutions; and the Convolutional Neural Operator (CNO)\cite{raonic2023convolutional}, which replaces them with local regular convolutions.
Despite their success, these models exhibit spectral bias \cite{rahaman2019spectral} - that is, they struggle to capture high-frequency components, which limits their accuracy for systems with rich multiscale dynamics, such as turbulence.
}


{
The architecture of the neural operator plays a critical role in how well it handles high-frequency content. 
U-shaped convolutional architectures, such as CNO and Time-Conditioned UNet (TC-UNet) \cite{ovadia2025real, gupta2022towards}, preserve the fine-scale details better than global architectures like FNO or DeepONet \cite{gupta2022towards, oommen2025integrating}. 
However, even these models struggle when applied to turbulent flows, where the energy spectrum follows a power-law decay across a wide range of wavenumbers. 
This difficulty becomes more pronounced at higher Reynolds numbers, where the inertial range extends deep into the high-frequency regime. 
In such cases, even expensive UNet-based operators often fail to model the dynamics associated with small-scale structures. In this work, we examine the root cause of this failure - not in the architecture, but in the way these models are trained.
}

{
Conventionally, neural operators are trained by minimizing a Euclidean loss, typically $L^1$ or $L^2$ error between the predicted and true states. 
While effective in reducing global error, this loss function implicitly favors low-frequency components, which dominate the energy spectrum in turbulent flows. 
As a result, the model can achieve low loss values while ignoring high-frequency content, leading to over-smooth predictions. 
Even architectures that are capable of representing high frequencies may fail to learn them under MSE-like training. 
Addressing this requires rethinking how we train neural operators."
}


{
Several recent works have targeted spectral bias in neural networks/operators by introducing clever techniques to penalize spectral content.
They include phase‑shift networks, multi‑scale DNNs, and adaptive activations to better balance learning across low and high frequency modes~\cite{xu2025understanding}. 
A similar multi‑scale branching strategy within the Fourier Neural Operator~\cite{you2024mscalefno} has also been investigated.
Chakraborty \emph{et al.} \cite{chakraborty2025binned} introduced the Binned Spectral Power loss, a frequency‑domain loss that bins spectral energy and adaptively weights errors across bands to explicitly penalize mismatches in energy distribution. 
Khodakarami \emph{et al.} introduced high-frequency scaling (HFS)\cite{khodakarami2025mitigating}, where they  decomposed the latent feature maps into non-overlapping patches, separating the low and high-frequency components, which helped them achieve more accurate and localized representations of multiscale flows. 
Physics-informed neural networks (PINN) \cite{raissi2019physics} also suffer from spectral bias \cite{toscano2025pinns} because, again, the $L^2$ norm of the PDE residuals is typically minimized. 
Recent remedies include Fourier feature embeddings \cite{wang2021eigenvector}, architectural modifications \cite{wang2024piratenets}, and improved optimizers \cite{wang2025gradient, wang2025simulating}.
However, in practice, PINNs must be retrained for each new initial or partially observed condition, making it less practical for real-time inference.
There have also been efforts to leverage recent advances in generative models for modeling multiscale systems like turbulence.
}


{ 
At its core, generative models are trained to synthesize new samples that fall within the training distribution, often conditioned on partial or low-fidelity information. 
By learning the full probabilistic structure of the data, they naturally mitigate the low‑frequency bias that burdens deterministic operators.
Generative adversarial networks (GANs)\cite{goodfellow2014generative}, diffusion models\cite{ho2020denoising}, and, more recently, flow matching \cite{lipman2022flow} techniques have all proven effective at learning turbulent flow dynamics.
Yousif \emph{et al.}\cite{yousif2022super} used a GAN to spatially super-resolve slices of the instantaneous velocity field in channel flow with longitudinal ribs from their downsampled counterparts.
Li \emph{et al.}\cite{li2023generative} employed a GAN to infer one instantaneous velocity component from another in rotating turbulent flow, and achieved closer statistical alignment with DNS than CNNs and EPOD.
GAN-based workflows have also been used for spatially super-resolving isotropic\cite{nista2024influence}, wall\cite{wu2024high, cuellar2024three}, and reactive turbulent flows\cite{bode2021using}.
Similarly, diffusion model-based frameworks \cite{shu2023physics, sardar2024spectrally} have been applied to two-dimensional Kolmogorov flows.
Although the diffusion models can estimate high-resolution states with good spectral fidelity, the iterative denoising makes them more expensive compared to GANs.
So it would be useful to evaluate the trade-off between inference cost and spectral fidelity across different generative methods.
Moreover, many of these studies focus on spatial super-resolution, whereas temporal super-resolution is also vital in turbulence, especially in settings like Schlieren systems, where achieving higher frame rates requires expensive high-speed cameras.
}

{
To simulate the temporal evolution of turbulent flows, Du \emph{et al.} \cite{du2024conditional} introduced the Conditional Neural Field Latent Diffusion (CoNFiLD) model, which integrates conditional neural-field encoding with latent diffusion to generatively synthesize spatiotemporal turbulence. 
While CoNFiLD produces time-coherent trajectories with strong spectral and statistical fidelity, its inference relies on iterative latent-diffusion sampling and is not formulated as a supervised, direct next-state forecaster mapping $u_t \rightarrow u_{t+\Delta t}$.
Oommen \emph{et al.} \cite{oommen2025integrating} trained a NO to directly forecast the future states, and used a diffusion model conditioned on the NO output to recover the fine-scale turbulent features that were smeared out due to the spectral bias of NO.
Molinaro \emph{et al.} \cite{molinaro2024generative} introduced GenCFD by directly employing a diffusion model conditioned on the initial condition and physical lead time to generate time-evolving turbulent flow fields, achieving improved mean‑field reconstruction and energy‑spectrum alignment on several test cases compared to standard NOs.
However, during inference, each time step of the generated trajectory requires iterative denoising-based sampling using the diffusion model. 
While this is faster than DNS and can be further accelerated by batch-based parallelization in modern GPUs, avoiding iterative denoising is still desirable. 
}

{
The generative models discussed above learn the functional prior of turbulent flows by training on ensembles of simulated trajectories. 
During inference, they can sample from the posterior distribution of flow fields conditioned on available observations.
However, in practice, previous works like CoNFiLD and GenCFD require on the order of $10^2$ to $10^3$ volumetric training trajectories to effectively learn these priors.
In turbulence, generating $10^2$ to $10^3$ volumetric trajectories using high-fidelity DNS incurs a substantial computational burden, which displaces the practical advantage of faster downstream inference.
Leveraging NOs with stronger inductive biases, enhanced with generative modeling principles to mitigate spectral bias, can make turbulence surrogates substantially more data efficient.
}

{
Beyond forecasting, generative models have also been adopted for reconstructing full turbulent flow fields from sparse measurements.
Parikh \emph{et al.} \cite{hemant2025conditional} used flow matching to reconstruct full turbulent fields using only sparse wall measurements in near-wall turbulence.
Similarly, Huang \emph{et al.} \cite{huang2024diffusionpde} and Buzzy \emph{et al.} \cite{buzzy2025polymicros} demonstrated that diffusion models conditioned on partial measurements can reconstruct full functions. 
Chakraborty \emph{et al.}\cite{chakraborty2025multimodal} also used a conditional diffusion model to reconstruct ERA5 atmospheric data from spatially-unstructured IGRA radiosonde observations.
These advances suggest a natural extension to experimental fluid mechanics, where full turbulent flow fields could be reconstructed from sparse PTV measurements, providing a bridge between limited experimental data and high-fidelity flow representations.
}

{
\vspace{3mm}
\noindent\textbf{Contributions:} Building on these developments, the major contributions of our work are as follows.
\begin{enumerate}
    \item We propose an adversarially trained neural operator (adv-NO) that bridges operator learning and generative modeling for spatio-temporal super-resolution and forecasting, leading to $10\times$ lower energy-spectrum error than vanilla NO and 114$\times$ speed-up compared to current generative model benchmarks.
    \item We adapt a conditional diffusion model for zero-shot flow reconstruction from sparse PTV-style velocity measurements and achieve phase alignment in the observed regions while recovering correct global statistics, a setting directly relevant to experimental fluid mechanics.
    \item We demonstrate both forecasting and flow reconstruction under severely limited data, training on only a single trajectory for each task, using 160 snapshots for forecasting and 150 for reconstruction.    
\end{enumerate}

Additionally, we perform detailed comparisons and analyses. 
(a) Compared to diffusion models, GANs, and VAEs, adv-NO offers a favorable balance between inference cost and spectral fidelity.
(b) Physics-informed (divergence-penalized) variants fail to mitigate spectral bias.
(c) Parseval-Plancherel analysis explains why conventional $L^2$-trained NOs underrepresent high frequencies (spectral bias).
(d) Fourier analysis of learned kernels reveals that adv-NO naturally develops a high-pass filtering behavior, mitigating spectral bias. 
A detailed outline of this study is illustrated in \autoref{fig:outline}. 
We present our results in the next section, followed by the Discussion and Methods sections.

}

{
\begin{figure}[h!]
  \centering
  \includegraphics[width=0.98\linewidth]{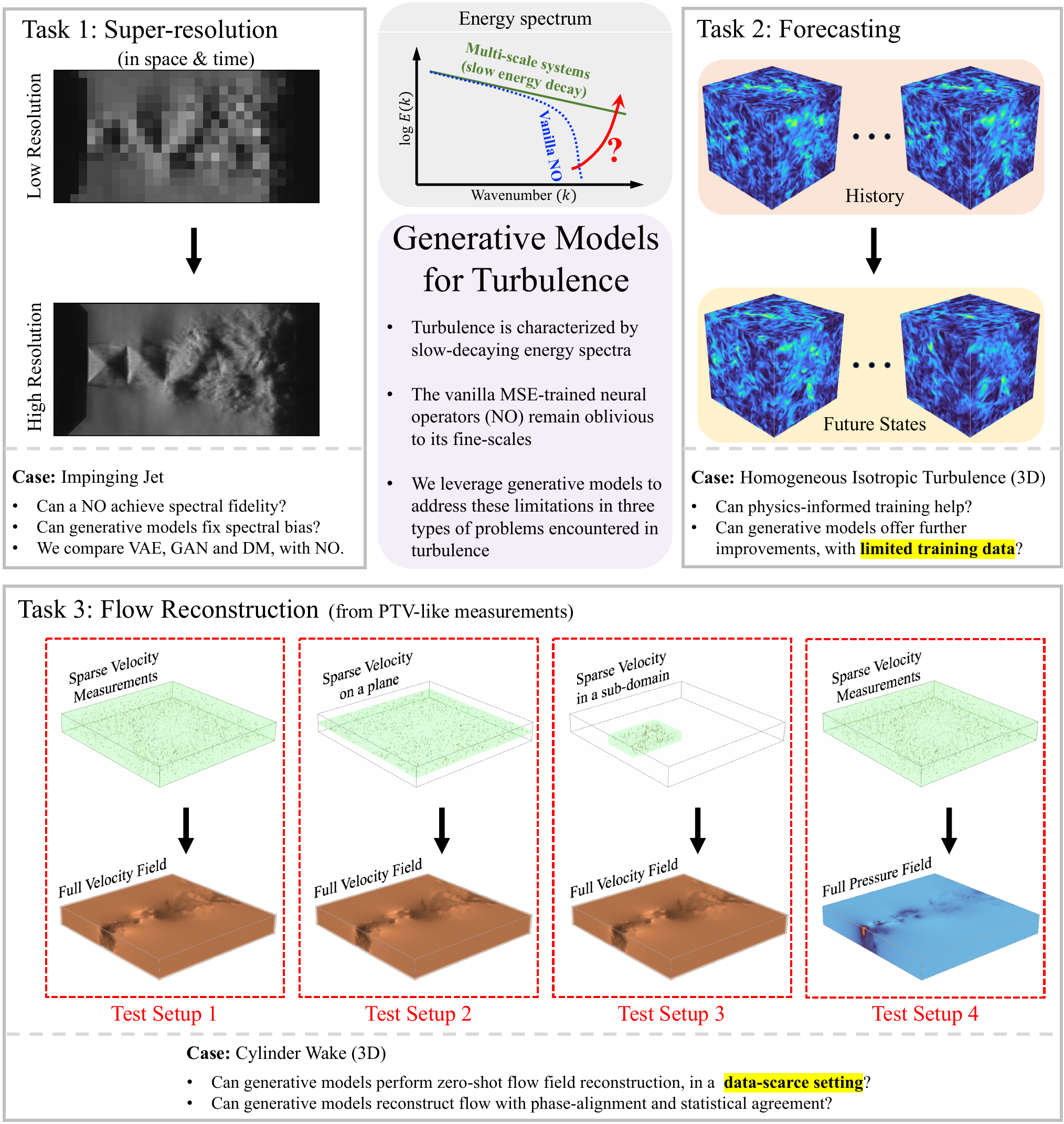}
  \caption{\textbf{Outline of the study.} Schematic overview of the three turbulence modeling tasks investigated in this work: 
  (1) super-resolution of low-resolution flow fields in both space and time, 
  (2) forecasting turbulent flow evolution from limited training data, and 
  (3) zero-shot full-field reconstruction from partial observations. 
  We compare conventional MSE-trained neural operators (NO) against generative models and physics-informed variants, highlighting mitigation of spectral bias.
  For 3D tasks, we train with limited data, acknowledging the difficulty of obtaining high-fidelity DNS or experimental datasets.
  }
  \label{fig:outline}
\end{figure}
}

\section*{Results}
\label{sec:results}

\subsection*{Spatio-temporal super-resolution of Schlieren visualizations}

{
High-speed Schlieren imaging is widely used to visualize compressible flows \cite{settles2017review, settles2022schlieren}.
Here, we consider impinging jet experiments to capture the interaction between a compressible air jet and a flat plate using Schlieren imaging, which visualizes gradients in fluid density. 
The flow is under-expanded with a nozzle exit Mach number of 1, and the jet is directed at ambient temperature onto the surface. A standard Z-type Schlieren setup is used to record the resulting shock structures and flow features.
Further details can be found in \cite{wang2022deep, zhang2025operator}.
Here, we train our surrogate NO to perform spatio-temporal super-resolution, reconstructing high-resolution, high-framerate (HRHF) states from low-resolution, low-framerate (LRLF) Schlieren visualizations. 
Unlike typical computer vision super-resolution, which focuses solely on spatial upscaling, we simultaneously recover fine spatial details and intermediate temporal states.
}

{
The dataset consists of 1000 snapshots with a resolution of $[128,256]$ spanning the spatial domain $[-0.8, 5.6] \times [-1.45, 1.75]$.
Each snapshot is separated by a time interval of $\tau = 4.76 \mu s$.
The first 800 timesteps are used for training, the next 100 for validation and the last 100 for testing. 
The original dataset is the HRHF representation ($u_{HRHF}(\bm{x},t)$). 
The LRLF state ($u_{LRLF}(\bm{x},t)$) is created by subsampling the HRHF by a factor of 8 along space and 4 along time as shown in \autoref{fig:genai_comparison} a). 
We train the surrogate NO, TC-UNet\cite{ovadia2025real}, to learn the mapping from $[u_{LRLF}(\bm{x}, t+\tau),u_{LRLF}(\bm{x}, t+5\tau)] \rightarrow [u_{HRHF}(\bm{x}, t+\tau),u_{HRHF}(\bm{x}, t+2\tau),u_{HRHF}(\bm{x}, t+3\tau),u_{HRHF}(\bm{x}, t+4\tau),u_{HRHF}(\bm{x}, t+5\tau)]$.
}

{
As discussed earlier, the conventional NO trained to minimize the Euclidean error suffers from spectral bias and subsequently estimates over-smoothed states. 
We compare predictions of NO with other generative algorithms often used in the computer vision community for image super-resolution problems. 
Specifically, we train variational autoencoder (VAE), generative adversarial network (GAN), and diffusion model (DM) to super-resolve the over-smoothed NO predictions (NO+Gen) as shown in \autoref{fig:genai_comparison} b).
We also consider an NO that was adversarially trained against a discriminator (adv-NO).  
In \autoref{fig:genai_comparison} c) and d), we compare the energy spectra and predicted density gradients, respectively, at $t+3\tau$.
\autoref{fig:genai_comparison} e), compares the error in the log of energy spectrum against the per-sample inference cost in terms of giga-floating-point operations (GFLOPs).
Additionally, we report the normalized-root-mean-squared-error (nRMSE) of the field and log-energy-spectrum, number of trainable parameters, and compute cost in terms of per-sample GFLOPs, inference time and peak VRAM usage in \autoref{tab:cost_accuracy_sr}.
}

{
\begin{figure}[h!]
  \centering
  \includegraphics[width=0.98\linewidth]{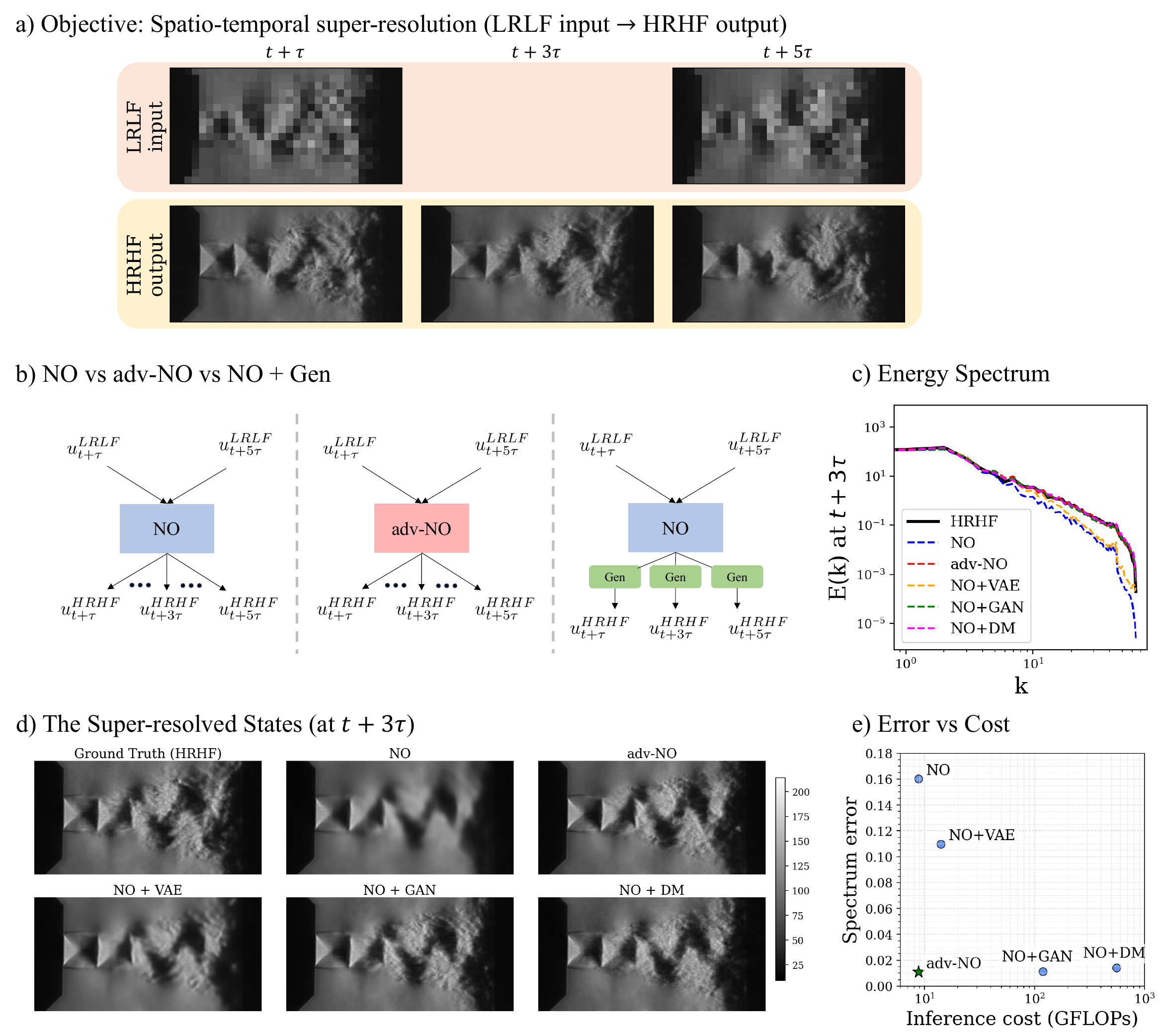}
  \caption{\textbf{Spatio-temporal super-resolution.}
    a) Our objective: learn the mapping from low resolution low framerate (LRLF) input to high resolution high framerate (HRHF) output.
    b) Schematic of learning setups for conventionally trained neural operator (NO), adversarially trained NO (adv-NO), and NO combined with generative (Gen) models - variational autoencoder (VAE), generative adversarial network (GAN), and diffusion model (DM). Gen is used to super-resolve NO-predicted states.
    c) Comparing the energy spectrum at $t+3\tau$.
    d) Super-resolved density gradients predicted by NO, adv-NO, and NO+Gen models.
    e) Spectrum error versus inference cost per sample for five models. Adv-NO attains low error at orders-of-magnitude lower computational cost than NO+GAN and NO+DM while outperforming NO and NO+VAE with similar costs. 
    }

  \label{fig:genai_comparison}
\end{figure}
}

{
\begin{table}[t]
\centering
\small
\setlength{\tabcolsep}{6pt}
\begin{tabular}{
l
S[table-format=1.4]
S[table-format=1.4]
c
S[table-format=3.2]
S[table-format=1.3]
S[table-format=1.4]
}
\toprule
\textbf{Model} & \multicolumn{1}{c}{\makecell{\textbf{Field Error}\\\textbf{(NRMSE)}}} & \multicolumn{1}{c}{\makecell{\textbf{Energy-Spectrum Error}\\\textbf{(NRMSE)}}} & \multicolumn{1}{c}{\makecell{\textbf{\# params}\\\textbf{(M)}}} & \multicolumn{1}{c}{\makecell{\textbf{Compute}\\\textbf{(GFLOPs)}}} & \multicolumn{1}{c}{\makecell{\textbf{Inference time}\\\textbf{(s)}}} & \multicolumn{1}{c}{\makecell{\textbf{Peak VRAM}\\\textbf{(GB)}}} \\
\midrule
NO       & 0.0542 & 0.1601 & 2.4            & 8.82   & 0.010 & 0.2898 \\
adv-NO   & 0.0662 & 0.0109 & 2.4            & 8.82   & 0.010 & 0.2898 \\
NO+VAE   & 0.0615 & 0.1098 & 2.4 + 2.1    & 14.08  & 0.014 & 0.2982 \\
NO+GAN   & 0.0646 & 0.0112 & 2.4 + 1.7    & 118.92 & 0.014 & 0.2964 \\
NO+DM    & 0.0692 & 0.0140 & 2.4 + 2.4    & 558.78 & 0.697 & 0.3171 \\
\bottomrule
\end{tabular}
\caption{
Errors and computational costs for models in the super-resolution study. 
Compute, inference time, and peak VRAM were measured for one sample on the same device and precision.
}
\label{tab:cost_accuracy_sr}
\end{table}
}

{
Adv-NO, NO+GAN, and NO+DM achieve the best energy spectrum, followed by NO+VAE.
Adv-NO offers the best accuracy-compute trade-off, recovering spectra and sharp gradients at NO-like inference cost.
In \autoref{fig:genai_comparison} e), NO+GAN appears more computationally demanding because its generator super-resolves NO outputs using residual-in-residual dense blocks (RRDB) \cite{wang2018esrgan}, which incur higher compute than conventional residual blocks. 
Similarly, NO+DM performs iterative denoising-based super-resolution of NO predictions, adding 32 forward passes per sample, and thus further increases compute.
While adv-NO and NO+Gen recover high-frequency content, the fine scales of their predictions are not perfectly phase-aligned with the ground truth, yielding slightly higher field errors than NO\cite{oommen2025integrating}, as shown in \autoref{tab:cost_accuracy_sr}.
We also compare super-resolved states estimated using NO and adv-NO with bicubic interpolation in \autoref{sec:imp_jet_bicubic}.
While these results demonstrate how generative models can improve spatial and temporal fidelity within known sequences, we now turn to the harder challenge of forecasting turbulent flow evolution beyond the observed training window.
}

\subsection*{Forecasting with limited training data}

{
Training a data‐driven surrogate to forecast turbulent flows is notoriously difficult because generating high‐fidelity volumetric, time‐resolved ground truth is extremely expensive, and experimental measurements of full 3D fields are rarely available. 
In practice, one is often limited to a handful of snapshots from a direct numerical simulation. 
With that constraint, we train our neural operator to forecast three‐dimensional forced homogeneous isotropic turbulence (HIT) using only 160 time‐steps from a single simulated trajectory for training.
}

{
HIT is a classical benchmark because, unlike shear dominated flows, it isolates the non-linear energy cascade of small‐scale eddies under statistically stationary conditions. 
We generated our HIT dataset with a spectral‐element solver nekRS (\cite{NEKRS})
at a Taylor Reynolds number $Re_{\lambda} = 90$, where $Re_{\lambda} = u_{rms}\lambda/\nu$ characterizes the ratio of inertial to viscous scales ($\lambda$ is the Taylor microscale, $\nu$ the kinematic viscosity). 
The flow state was saved every $t = \tau (= 0.04)$ (normalized time units). 
The eddy-turnover time \cite{ruetsch1992evolution} $t_E$ was computed as $u_{rms}^2/\varepsilon$, where $\varepsilon = 2\nu S:S$ is the mean dissipation rate and $S = \frac{1}{2}(\nabla \mathbf{u} + \nabla \mathbf{u}^T)$ is the rate‐of‐strain tensor. 
Using this single 160‐step trajectory, we try to forecast five eddy-turnover time scales (see Training and Implementation Details in the Methods section for architecture sizes and training configurations).
To effectively train the surrogates to learn the evolution dynamics under limited training data, we split the available single trajectory of 160 snapshots of $\bm{u}=[u,v,w,P]$ to multiple sub-trajectories.
Specifically, the surrogate NO and its variants are trained to learn the mapping - $[\bm{u}(\bm{x},t-3\tau), \cdots, \bm{u}(\bm{x},t)] \rightarrow [\bm{u}(\bm{x},t+\tau), \cdots, \bm{u}(\bm{x},t+4\tau)]$.
During inference, the states beyond $t+4\tau$ are estimated autoregressively, by feeding the surrogate model's prediction as its own input.
In this experiment, we compare the standard NO, adv-NO, and a physics-informed variant of the NO that is constrained to satisfy the mass-conservation by penalizing the divergence of the velocity field vector computed using spectral derivatives.
We call it $\text{NO}+\nabla\cdot u$ in this study.
The DNS was performed using the spectral element method\cite{moxey2020nektar++}, which satisfies the conservation laws at the quadrature points of each spectral element in a weak sense.
For training NO and its variants, the DNS solution was interpolated using the solver's polynomial basis from quadrature points onto an equi-spaced grid, where the divergence-free property is no longer satisfied.
To address this issue, we enforce incompressibility by solving for $\phi$ from $\nabla^2 \phi = \nabla\cdot \bm{u}$, and using $\bm{u} - \nabla \phi$ as the updated solution.
The results of this study are provided in \autoref{fig:hit3d_ke}.
}

{
\begin{figure}[h!]
  \centering
  \includegraphics[width=0.95\linewidth]{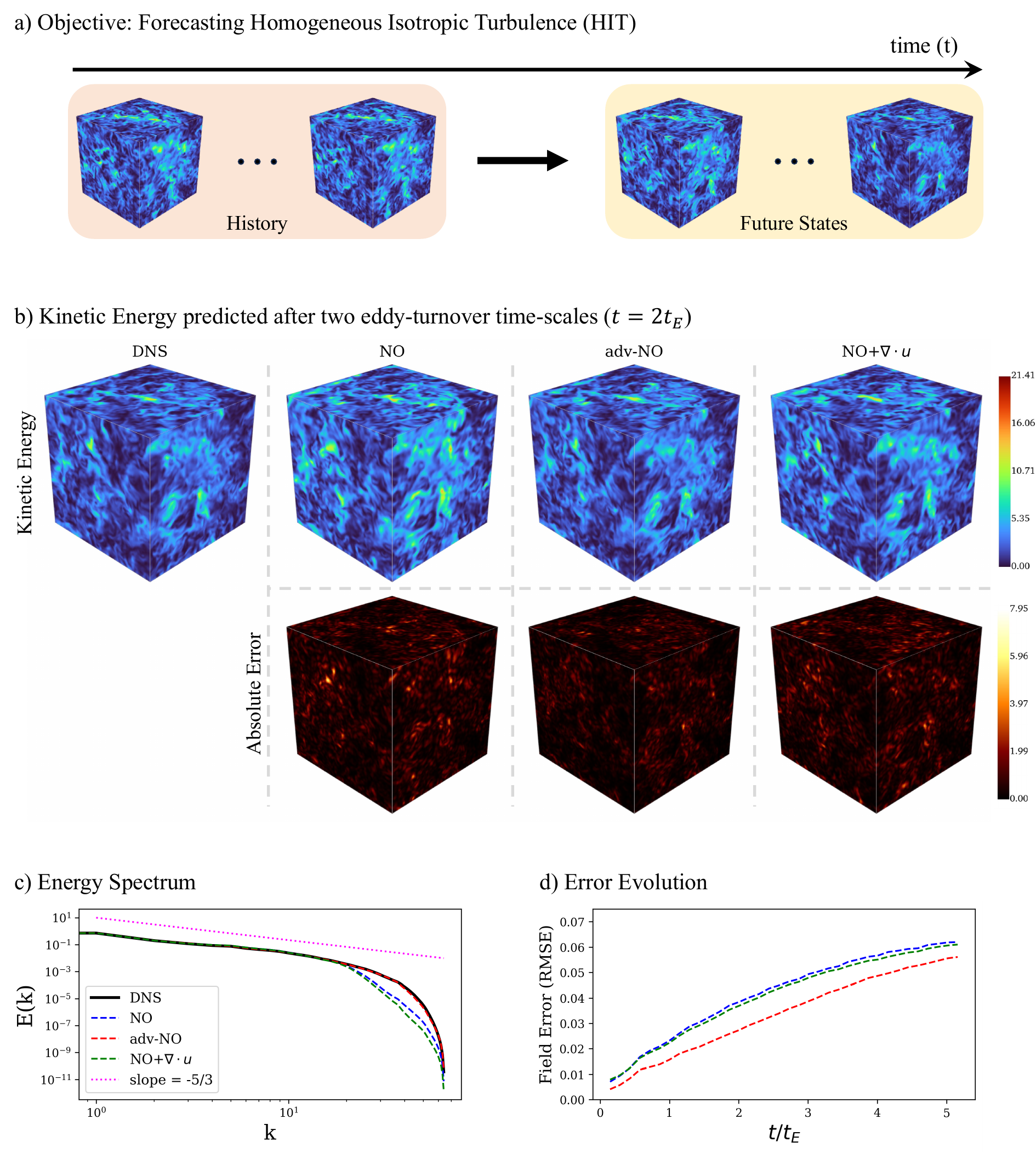}
  \caption{\textbf{Forecasting Homogeneous Isotropic Turbulence ($Re_{\lambda}=90$).}
  The kinetic energy (KE) estimated using the spectral element method is used as a reference for comparing the predictions of NO, adv-NO, and a physics-informed variant of the neural operator that penalizes the divergence of the velocity vector.
  All the models were trained only on 160 timesteps of a single trajectory.
  The states at $t=2t_E$ ($t_E$ is the eddy-turnover time) are visualized in b). 
  The energy spectrum averaged across time is provided in c). 
  The evolution of RMSE of $\bm{u}$ over the entire forecast horizon ($5t_E$), is provided in d).
  }
  \label{fig:hit3d_ke}
\end{figure}
}

{
We compare the kinetic energy (KE) at $\Delta t=2t_E(=14\tau)$ predicted by NO, adv-NO and NO+$\nabla\cdot u$ against the reference DNS simulated using the spectral element method in \autoref{fig:hit3d_ke} b).
The velocity and pressure states are separately visualized in \autoref{fig:hit3d_uvwP}.
The KE and the $[u,v,w,P]$ states predicted by adv-NO achieve better phase alignment with DNS compared to NO and NO+$\nabla\cdot u$.
The time-averaged energy spectrum compared in \autoref{fig:hit3d_ke} c) further highlights the over-smoothing/spectral bias suffered by NO.
The energy spectra comparison between NO and NO+$\nabla\cdot u$ suggests that penalizing the divergence does not correct the spectral bias of a conventional $L^2$‐trained NO.
This result is explainable because NO+$\nabla\cdot u$ also minimizes an $L^2$ norm of the mass-conservation PDE residual, and therefore, struggles to minimize the fine-scale features that arise in the PDE residual of the NO prediction due to spectral bias.
That being said, enforcing incompressibility does help NO+$\nabla\cdot u$ achieve a better phase-alignment than NO.
In \autoref{fig:hit3d_ke} d), we investigate how the field error evolves across the entire forecast horizon of over five eddy-turnover time scales (which corresponds to 36 $\tau$).
For the field error, we take the RMSE of the normalized velocity and pressure states at each timestep.
The time-evolution of kinetic energy and energy spectrum is further visualized in \autoref{fig:hit3d_ke_time}.
The adv-NO is able to preserve the Kolmogorov scaling throughout the forecast horizon and maintains phase alignment well until $4t_E$.
}

{
We also benchmark adv-NO against GenCFD~\cite{molinaro2024generative}, a conditional score-based diffusion model that conditions on the ``lead time" $\Delta t$ and generates forecasts through iterative denoising. 
In our low-data setting, a one-shot GenCFD mapping from $\bm{u}(\mathbf{x},t)$ to $[\bm{u}(\mathbf{x},t{+}\tau),\ldots,\bm{u}(\mathbf{x},t{+}36\tau)]$ did not converge because of the weak input-output correlation between states separated by a $\Delta t = 36\tau (\sim5\,t_E)$. 
We therefore adopt the same $4\tau$-ahead setup used for adv-NO and roll out autoregressively to $5\,t_E$. 
Both models have $\approx 6$M parameters and are trained for $48$ GPU-hours. 
Except for the above-mentioned changes, the default training/inference settings were adopted unchanged. 
GenCFD is unable to reconstruct the fine-scale turbulent structures and outputs noisy predictions under the data-scarce setting, as shown in \autoref{fig:gencfd_comparison}.
Forecasting 36 timesteps with GenCFD takes 228s on an NVIDIA-H100 GPU, because of the iterative denoising of each forecasted timestep, making it over a hundred times slower than adv-NO.
A detailed error-cost analysis is provided in \autoref{tab:cost_accuracy_forecasting}.
}

{
\begin{figure}[h!]
  \centering
  \includegraphics[width=0.99\linewidth]{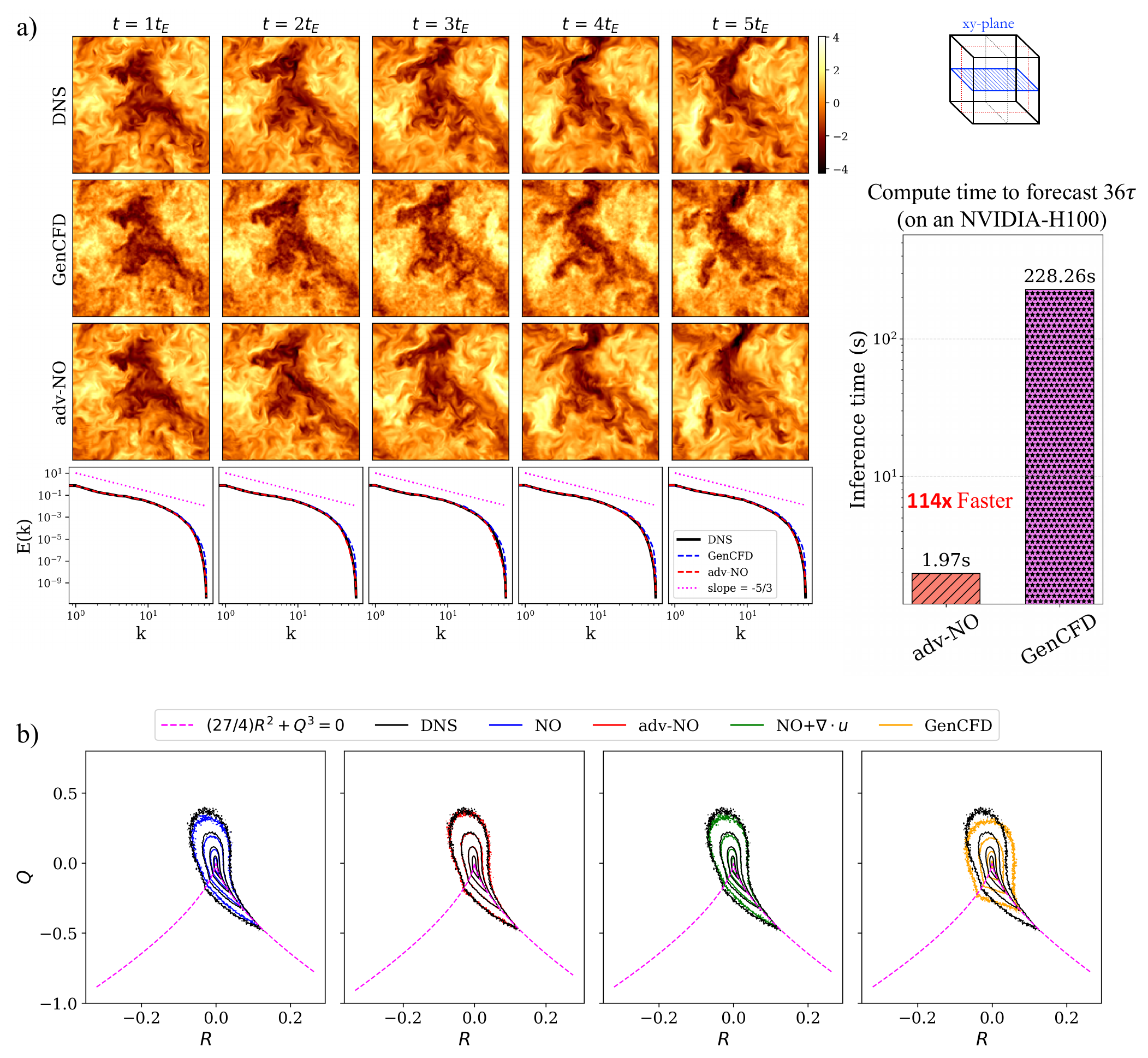}
  \caption{\textbf{Comparison with diffusion model baseline.}
  a) We compare adv-NO with GenCFD for forecasting homogeneous isotropic turbulence. 
  The mid-$xy$ plane of the $u$-velocity is visualized. 
  In this data-scarce setting, GenCFD predictions appear noisy and struggle to reconstruct fine-scale structures, whereas adv-NO retains small-scale detail without requiring iterative denoising. 
  Inference time to forecast $36\,\tau$ on an H100 is also reported. 
  Keeping the model size and training costs the same, adv-NO is $\sim114\times$ faster than GenCFD.
  b) We plot and compare the isocontours of the joint probability distribution of $Q$ and $R$ at $t=5t_E$. 
  Adv-NO aligns best with DNS. 
  }
  \label{fig:gencfd_comparison}
\end{figure}
}

{
We further assess small-scale structure via the invariants of the velocity-gradient tensor \(A=\nabla\mathbf{u}\), defined as the coefficients of its characteristic polynomial
\(\lambda^{3}+P\lambda^{2}+Q\lambda+R=0\), with \(P=-\mathrm{tr}\,A\) (\(P=0\) for incompressible flow), \(Q=-\tfrac12\mathrm{tr}(A^{2})\), and \(R=-\det A\).
Writing \(A=S+\Omega\) with strain \(S=(A{+}A^{\!\top})/2\) and rotation \(\Omega=(A{-}A^{\!\top})/2\), one has
\(Q=\tfrac12(\|\Omega\|_{F}^{2}-\|S\|_{F}^{2})\), so \(Q>0\) indicates rotation-dominated regions and \(Q<0\) strain-dominated regions.
We normalize \(Q\) by \(\langle A_{ij}A_{ij}\rangle\) and \(R\) by \(\langle A_{ij}A_{ij}\rangle^{3/2}\).
We compare the joint PDF \(p(Q,R)\) by overlaying isocontours (Fig.~\ref{fig:gencfd_comparison}\textbf{b}), which exhibits the typical tear-drop shape in the \(Q\)-\(R\) plane.
Adv-NO aligns best with the DNS.
}

{
\begin{table}[t]
\centering
\small
\setlength{\tabcolsep}{6pt}
\begin{tabular}{
l
S[table-format=1.4]
S[table-format=1.4]
c
S[table-format=3.2]
S[table-format=1.3]
S[table-format=1.4]
}
\toprule
\textbf{Model} & \multicolumn{1}{c}{\makecell{\textbf{Field Error}\\\textbf{(NRMSE)}}} & \multicolumn{1}{c}{\makecell{\textbf{Energy-Spectrum Error}\\\textbf{(NRMSE)}}} & \multicolumn{1}{c}{\makecell{\textbf{\# params}\\\textbf{(M)}}} & \multicolumn{1}{c}{\makecell{\textbf{Compute}\\\textbf{(GFLOPs)}}} & \multicolumn{1}{c}{\makecell{\textbf{Inference time}\\\textbf{(s)}}} & \multicolumn{1}{c}{\makecell{\textbf{Peak VRAM}\\\textbf{(GB)}}} \\
\midrule
NO                   & 0.0326 & 0.0865 & 6.2 & 1056.38  & 0.082 & 1.8959  \\
adv-NO               & 0.0265 & 0.0235 & 6.2 & 1056.38  & 0.082 & 1.8959  \\
NO+$\nabla\cdot u$   & 0.0320 & 0.1262 & 6.2 & 1056.38  & 0.082 & 1.8959  \\
GenCFD               & 0.0327 & 0.0587 & 6.2 & 45297.99 & 7.804 & 28.5566 \\
\bottomrule
\end{tabular}
\caption{
Errors and computational costs for models in the forecasting study. 
Compute, inference time, and peak VRAM were measured for one sample on the same device and precision.
}
\label{tab:cost_accuracy_forecasting}
\end{table}
}

\subsection*{Flow reconstruction from partial observations}

{
Unlike forecasting, where we predict future flow states from a history of complete snapshots, many experimental scenarios involve the inverse problem of reconstructing the full flow field from only sparse and partial measurements.
In experimental fluid mechanics, the complete flow field is rarely accessible. 
Techniques such as particle tracking velocimetry (PTV) provide velocity measurements only at the locations of seeded particles, resulting in inherently sparse data. 
These measurements are typically sparse \cite{toscano2025aivt} and are often restricted to a smaller subdomain or a single plane of the flow\cite{kiran2025influence}. 
Direct pressure measurements are even more challenging. 
Here, we aim to reconstruct the full three-dimensional velocity and pressure fields from sparse PTV-like velocity observations, ensuring phase alignment near the measurement region while also matching the global flow statistics, under very limited number of training samples.
}

{
To achieve this, we train a conditional generative model to sample directly from the posterior distribution $p(\mathrm{full\ flow} \mid \mathrm{partial\ observations})$.
In this study, the model is trained on high-fidelity DNS data of the cylinder wake at $Re=\frac{U_{\infty} D}{\nu}=11,000$, where $U_{\infty}$ is the inflow velocity, $D$ is the diameter of the cylinder and $\nu$ is the kinematic viscosity, generated by using the spectral element code nekRS (\cite{NEKRS}) in the domain $[-12D,20D]\times[-10D,10D]\times[0,9.6D]$, where the cylinder axis is placed on $x=0, y=0$. Nonetheless, the training data is collected in the domain $[-4D,4D]\times[-4D,4D]\times[0,2D]$. 
During training, the model learns to infer missing flow information by conditioning on partially observed inputs.
}

{
Specifically, we generate a binary mask for each of the four physical variables $[u, v, w, P]$.
The entries are randomly set to zero (missing) or one (observed) according to a mask fraction sampled uniformly from $\mathcal{U}(0,1)$.
Half of each mini-batch uses independent random masking per channel, while the other half uses a single random mask replicated across all channels. 
This mask is multiplied with the flow field to produce the masked input, which we then concatenate with the mask itself. 
This concatenated tensor (with 4+4 channels) serves as the conditioning input to the model.
(See Algorithm \ref{alg:mask_prep_training} in SI).
Sampling mask fractions that range from 0 to 1, enables the training to span the full spectrum between deterministic reconstruction (mask fraction = 0, no missing data) and purely generative reconstruction (mask fraction = 1, all data missing).
This variability enables the model to get exposed to the entire continuum of partial observations.
}

{
We compare two conditional generative approaches: a diffusion model variant - EDM \cite{karras2022elucidating}, and a Real‑ESRGAN adapted for 3D flow reconstruction \cite{wang2021real}.
Here, we learn instantaneous flow field reconstruction from sparse measurements without learning the dynamics.
So we do not call our implementation adv-NO, but a GAN with a UNet generator that is not time-conditioned. 
See Methods for details on how we adapt these generative models, originally designed for 2D image synthesis, to the volumetric reconstruction of 3D vector fields.
Both models are conditioned identically on masked DNS data, allowing us to directly compare their ability to reconstruct full flow fields from sparse observations.
Training, validation, and test datasets consist of 150, 18, and 17 snapshots, respectively.
Again, our datasets are small to reflect the limited access to experiments and high-fidelity simulations. 
After training the models, we investigate the model's ability to zero-shot reconstruct the velocity fields from sparse velocity observations in the domain (Test Setup 1),  on a plane (Test Setup 2), in a sub-domain (Test Setup 3), and reconstruction of the full pressure field from sparse velocity observations (Test Setup 4), as shown in \autoref{fig:FR}.
For each Test Setup, the input velocity measurements are randomly sampled from the region highlighted in green, and the pressure field is completely masked out to mimic the real PTV experiments. 
}

{
\begin{figure}[h!]
  \centering
  \includegraphics[width=0.99\linewidth]{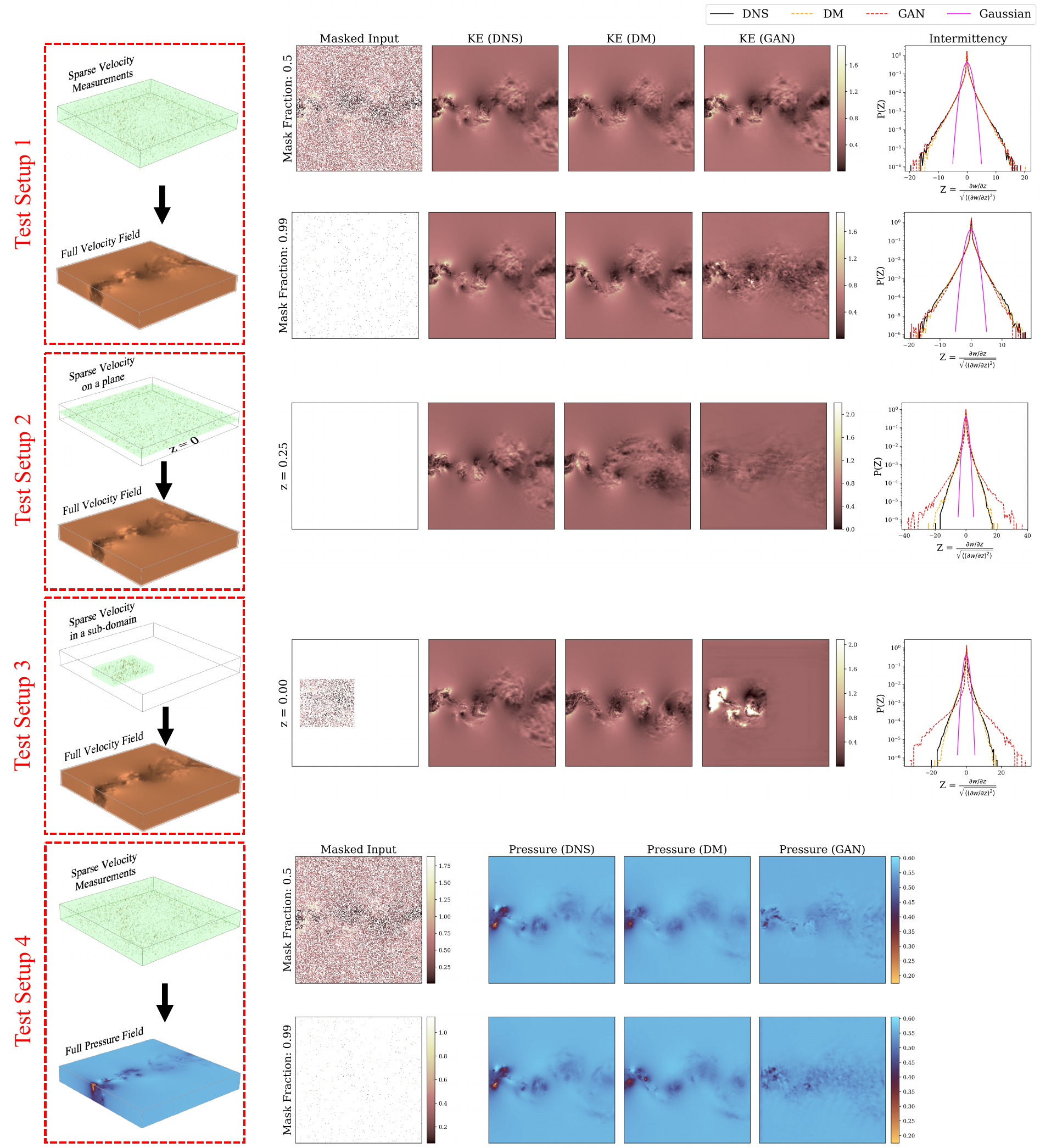}
  \caption{\textbf{Zero-shot Flow Reconstruction.}
  We evaluate zero-shot reconstruction from only sparse velocity measurements without requiring pressure inputs, thereby mimicking real PTV experiments. 
  Test setups 1 to 3 reconstruct the full velocity field from sparse velocity measurements. 
  Test Setup 1 - from measurements in the full domain (like volumetric PTV),
  Test Setup 2 - from measurements on a plane (like planar PTV),
  Test Setup 3 - from measurements in a sub-domain.
  In Test Setup 4, we reconstruct the full pressure field from sparse velocity measurements.
  We compare the PDFs of the gradient of the velocity (intermittency) reconstructed by GAN and DM (1 and 4 have the same kinetic energy). 
  All results are reported on the test dataset without further fine-tuning.}
  \label{fig:FR}
\end{figure}
}


{
In Test Setup 1, when the mask fraction is 0.5, both DM and GAN achieve good phase alignment with matching statistics. But GAN fails when the mask-fraction is increased to 0.99. 
The kinetic energy of the reconstructions, energy spectra, and intermittency for mask fractions = 0, 0.1, 0.3, 0.5, 0.7, 0.9, 0.99 are visualized in \autoref{fig:FR_TS1}. 
Both DM and GAN are able to successfully reconstruct the vortex street for all mask fractions (except GAN when mask fraction=0.99). 
A detailed illustration of Test Setup 2 is provided in \autoref{fig:FR_TS2}. 
The DM achieves strong phase alignment on the measurement plane ($z=0$). 
The phase alignment gradually decreases on planes that are farther from the observation plane. 
The corresponding energy spectra and intermittency indicate that the DM recovers turbulent statistics accurately.
In contrast, GAN severely overestimates the higher velocity gradients, as seen in the intermittency plots (\autoref{fig:FR}) and fails to reconstruct the vortex street. 
We observe a very similar trend for DM and GAN in the context of phase alignment and turbulent statistics, in Test Setup 3 (see \autoref{fig:FR_TS3}).
The pressure reconstructions corresponding to different mask fractions are illustrated in \autoref{fig:FR_TS4}.
The diffusion model achieves strong phase alignment for all the mask fractions considered. 
The GAN reconstructions are suboptimal for mask fractions up to 0.9, and fail when the mask fraction is 0.99.
Across all test setups, the conditional diffusion model demonstrates robust reconstruction capability from sparse observations, recovering both large and small-scale features even when the available data is severely limited.
}

\section*{Discussion}
\label{sec:discussion}

\subsubsection*{Model suitability across tasks}
{
In the flow reconstruction task, the GAN fails when there is a large missing region. 
We believe the issue arises from the UNet-based generator, whose local convolutions make it ineffective at aggregating global context when observations are extremely sparse \cite{zeng2022aggregated}.
In contrast, diffusion models learn to denoise from a wide range of data-corruptions/signal-to-noise ratios by learning score functions via Langevin dynamics, thereby enabling reconstruction of flow fields that are locally and globally consistent.   
For spatio-temporal super-resolution and forecasting, where extreme data sparsity is not an issue, adversarial training performs well and is, in fact, a better choice than diffusion models as it avoids the computational overhead of iterative denoising.
Next, we analyze why NOs struggle to learn high-frequency content by leveraging Parseval-Plancherel Identity\cite{kammler2007first}.
}

\subsubsection*{Spectral bias in NO}

{
All the cases investigated in the Results sections exhibit a slow-decaying energy spectrum, and we observe that the NO trained to minimize Euclidean errors like MSE predicts over-smoothed states.
The energy spectrum analysis further reveals that the energy of NO predictions at higher wavenumbers is significantly lower than the true reference. 
To better understand this behavior, we analyze the drawbacks associated with training a neural network/operator by minimizing the Euclidean Error, like the $L^2$ norm for instance. 
}

{
Let $f(x)$ be the true function approximated by the NO $f_{\theta}(x)$.
By the Parseval-Plancherel identity, which states that the total energy in the physical space equals that in spectral space, the squared $L^2$ norm of the error to be minimized can also be represented in the frequency domain as follows,
\begin{align}
    \|e(x)\|_2^2 &= \int_{\Omega} |f(x) - f_{\theta}(x)|^2 dx \\
    &= \int_{k_0}^{k_{max}} |\mathcal{F}(f)(k) - \mathcal{F}(f_{\theta})(k)|^2 dk,
\end{align}
where $\mathcal{F}(\cdot)$ represents Fourier transformation. 

Let $k_0<k_1<\cdots<k_{n-1}<k_{max}$ represent the wavenumbers used to separate the squared $L^2$ norm to $n$ segments.
The squared $L^2$ norm of the error can now be represented as,
\begin{align}
    \|e(x)\|_2^2 &= \int_{k_0}^{k_{1}} |\mathcal{F}(f)(k) - \mathcal{F}(f_{\theta})(k)|^2 dk +  \cdots + \int_{k_{n-1}}^{k_{max}} |\mathcal{F}(f)(k) - \mathcal{F}(f_{\theta})(k)|^2 dk \\
    &= \text{term}_1 +  \cdots + \text{term}_n.
\end{align}
}

{
When the system of interest exhibits a decaying energy spectrum, like turbulence, the system's energy is $E(k) = \int_{k_0}^{k_{max}} \mathcal{F}(f)(k) dk$, is such that $\text{term}_1 > \text{term}_2 > \cdots > \text{term}_n$.
For turbulent systems that are characterized by a power-law scaling of the energy spectrum, $\text{term}_1 + \text{term}_2 + \cdots + \text{term}_i$ could be orders of magnitude larger than $\text{term}_{i+1} + \cdots + \text{term}_n$ for several $1<i<n$.
Subsequently, a NO trained to minimize $\|e(x)\|^2$ prioritizes the lower wavenumbers that bear more energy than the higher wavenumbers that carry less energy.
}

{
Previous works have shown that gradient-based training prioritizes low frequencies before high ones-``spectral bias''~\cite{rahaman2019spectral, zhi2020frequency, sitzmann2020implicit, cai2024towards, wang2025spectral}. 
But these prior analyses mostly consider shallow networks and specific activations. 
For example, Xu \emph{et al.}\cite{zhi2020frequency} establish a per-parameter gradient ordering for a one-hidden-layer $\tanh$ network implying a rate ordering under gradient flow.
We instead work in the operator-learning regime, involving maps $\mathcal{N}:\mathcal{U}\to\mathcal{V}$ with physically realistic spectra, and models whose Fourier sensitivity decays with frequency at initialization.
Under these milder assumptions, we prove (i) the gradient mass at initialization concentrates on low wavenumbers; and (ii) trained solutions exhibit a quantifiable high-frequency energy deficit relative to the target. 
A detailed analysis is provided in \autoref{sec:sb_analysis}.
Our analysis is architecture-agnostic and yields a global integral bound on gradient mass together with an explicit tail-energy gap over realistic training horizons.
} 

{
Our analysis on the spectral bias of neural operators also implies that pointwise $L^{2}$ error is a poor error metric for evaluating multi-scale systems.
In this context, it overweighs low-wavenumber (high-energy) errors and underweighs high-wavenumber discrepancies, enabling smoothed predictions to achieve low errors. 
This bias is amplified further when a dominant mean/large-scale component is present (e.g., our impinging jet), where NO yields lowest NRMSE (see \autoref{tab:cost_accuracy_sr}) while erasing fine-scale structure completely. For this reason, we also include the error in energy-spectrum as an additional metric.
}
\newline

\subsubsection*{Fourier analysis of the learned kernels in NO and adv-NO}
{
To understand how adversarial training affects the NO, in \autoref{fig:weight_analysis} d), we compare the channel-wise mean of the Fourier transformations of convolutional kernels from the same TC-UNet-based NO architecture, trained in the conventional and adversarial way. 
For reference, in \autoref{fig:weight_analysis} a) and c) we show the FFT of a low-pass (Gaussian) filter and a high-pass (Laplace) filter.
For each of the encoder and decoder blocks, the conventional NO tends to filter out the high-wavenumber features more aggressively compared to the adv-NO. 
The Fourier analysis suggests that adversarial training of the NO facilitates easier transformation/transmission of the high-wavenumber information - a critical property for better modeling of turbulent systems. 
}

{
\begin{figure}[h!]
  \centering
  \includegraphics[width=0.98\linewidth]{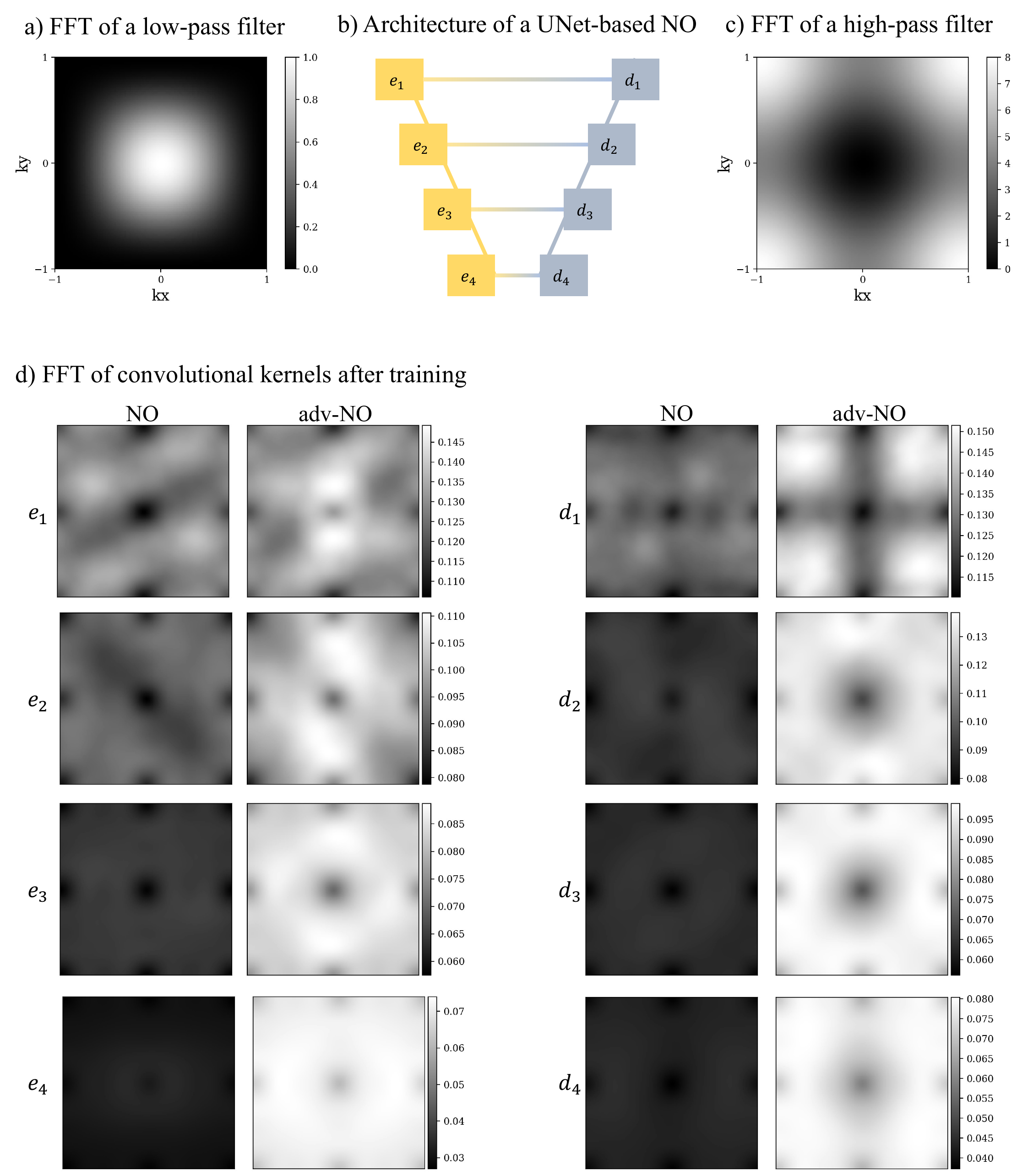}
  \caption{\textbf{Fourier analysis of the learned parameters.} a) and c) shows the Fourier transform of a low-pass (Gaussian) and high-pass (Laplace) filters, respectively.
  b) is a schematic of a UNet-based NO with 4 encoder ($e_i$) and 4 decoder ($d_i$) convolutional layers.
  d) Comparing the mean of the FFT magnitude of the learned convolutional kernels trained conventionally and adversarially. 
  The adv-NO kernels are more similar to the high-pass filter than NO kernels.
    }
  \label{fig:weight_analysis}
\end{figure}
}

\subsubsection*{Future Works}

{
The framework we propose is not limited to turbulence but can be applied out-of-the-box to a broad class of multiscale systems where data generation or collection is extremely challenging.
While our current work focuses on flows over regular grids in simple domains, Du et al. \cite{du2024conditional} demonstrated that generative models can seamlessly handle complex geometries with irregular meshes, albeit requiring large amounts of training data. 
Developing a framework that can handle complicated geometries while being data-efficient could be challenging but impactful.
In this work, we employed a UNet-based backbone for implementing both the GAN and diffusion models, which, while effective, may be less computationally scalable than transformer-based architectures. 
However, transformer-based frameworks often overfit rapidly in data-scarce settings.
Developing computationally scalable architectures that remain robust under limited-data conditions can be a significant advancement in the context of similar studies.
Leveraging physics-informed learning to extend our generative modeling framework to unseen flow configurations, boundary conditions, and geometries, without relying on additional simulations, presents an exciting new direction for future research. 
}

\section*{Methods}
\label{sec:methods}

\subsection*{Neural Operators (NO)}

Neural operators $\mathcal{G}$ are built to learn mappings between function spaces, with the goal of approximating the true underlying operator $\mathcal{N}$. 
Formally, for an operator $\mathcal{N} : \mathcal{U} \to \mathcal{V}$ defined between Banach spaces $\mathcal{U}$ and $\mathcal{V}$, we train a surrogate operator $\mathcal{G}_\theta$ to learn the mapping $\mathcal{U}\to\mathcal{V}$. 
Specifically, in super-resolution we learn $\mathcal{G}_\theta^{\mathrm{SR}}:\mathcal{U}^{\mathrm{LRLF}}\to\mathcal{U}^{\mathrm{HRHF}}$ that maps a sparsely sampled/low-resolution field $u_{\mathrm{in}}=u^{\mathrm{LRLF}}\in\mathcal{U}^{\mathrm{LRLF}}$ (defined only on a subset of $(x,y,t)$) to its dense/high-resolution counterpart $u_{\mathrm{tar}}=u^{\mathrm{HRHF}}\in\mathcal{U}^{\mathrm{HRHF}}$. 
In forecasting we learn $\mathcal{G}_\theta^{\mathrm{FC}}$ that advances the state in time, taking as input $u_{\mathrm{in}}=u(\cdot,t)$ or, more generally, a short history window 
$u_{\mathrm{in}}=[u(\cdot,t-(m-1)\Delta t),\ldots,u(\cdot,t)]$ and predicting the future field $u_{\mathrm{tar}}=u(\cdot,t+\Delta t)$. 
Both tasks are trained by supervised loss minimization over pairs $(u_{\mathrm{in}},u_{\mathrm{tar}})$:
\begin{equation}
    \theta^{*}
    = \arg\min_{\theta}\;
    \mathbb{E}_{(u_{\mathrm{in}},\,u_{\mathrm{tar}})}
    \left[ \left\|\,u_{\mathrm{tar}} - \mathcal{G}_{\theta}(u_{\mathrm{in}})\,\right\|_{p} \right],
    \label{eq:NO_obj}
\end{equation}
where for super-resolution $(u_{\mathrm{in}},u_{\mathrm{tar}})=(u^{\mathrm{LRLF}},\,u^{\mathrm{HRHF}})$ and for forecasting $(u_{\mathrm{in}},u_{\mathrm{tar}})=(u(\cdot,t\ \text{or history}),\,u(\cdot,t+\Delta t))$.
$\lVert \cdot \rVert_p$ denotes an appropriate $L^p$ norm ($p$ is typically 1 or 2), which measures the discrepancy between the target $u_{\mathrm{tar}}$ and the neural operator prediction $\mathcal{G}_\theta(u_{\mathrm{in}})$.

Here, we use a time-conditioned UNet \cite{ovadia2025real, gupta2022towards} as our neural operator, which takes $u_{\mathrm{in}}$ and $\Delta t$ as the input, and predicts $u_{\mathrm{tar}}$. 
This makes the operator continuous in time, while the spatial side is still discretized on a grid. 
Our model is mesh-dependent, but it still qualifies as operator learning since the target is the underlying operator $\mathcal{N}$.
The lack of strict mesh-invariance could raise concerns, but in practice these architectures often outperform Fourier Neural Operator (FNO) and its variants \cite{ovadia2025real, gupta2022towards}. 
One reason is the favorable inductive bias: convolutions naturally capture locality and translation symmetry in the data, and temporal conditioning makes the dynamics easier to model. 
This also makes the time-conditioned UNet especially effective in low-data regimes, which is often the case in turbulence, where DNS and experiments are expensive to generate/access.

\subsection*{Adversarially Trained Neural Operators (adv-NO)}

The $L^{p}$ objective in \autoref{eq:NO_obj} yields accurate low-wavenumber structure but under-penalizes errors at high wavenumbers, leading to over-smoothing (spectral bias) in both super-resolution and forecasting of multi-scale systems like turbulence. 
We resolve this limitation while retaining the operator’s temporal conditioning and fast inference by training the same $\mathcal{G}_{\theta}$ but adversarially against a discriminator.

\noindent Given an input $u_{\mathrm{in}}$ (LRLF snapshot for super-resolution or a history window for forecasting) and the target $u_{\mathrm{tar}}$, let
\[
u_{f} \;=\; \mathcal{G}_{\theta}(u_{\mathrm{in}}) 
\]
Since $u_f$ is estimated by the NO, we treat it as a ``fake'' state and $u_{\mathrm{tar}}$ as a ``real'' state. 
We adopt the relativistic average GAN (RaGAN) objective \cite{jolicoeur2018relativistic}, where a UNet-based discriminator estimates the relative realness between the real ($u_{\mathrm{tar}}$) and fake ($u_f$) samples as shown below.
\begin{equation}
L_{D} = - \mathbb{E}_{u_{\mathrm{tar}}} \left[ \log \sigma\!\big(D_{\psi}(u_{\mathrm{tar}}) - \mathbb{E}[D_{\psi}(u_f)]\big) \right] 
        - \mathbb{E}_{u_f} \left[ \log \big(1 - \sigma(D_{\psi}(u_f) - \mathbb{E}[D_{\psi}(u_{\mathrm{tar}})])\big) \right], 
\end{equation}
\begin{equation}
L_{G} = - \mathbb{E}_{u_{\mathrm{tar}}} \left[ \log \big(1 - \sigma(D_{\psi}(u_{\mathrm{tar}}) - \mathbb{E}[D_{\psi}(u_f)])\big) \right] 
        - \mathbb{E}_{u_f} \left[ \log \sigma(D_{\psi}(u_f) - \mathbb{E}[D_{\psi}(u_{\mathrm{tar}})]) \right],
\label{eq:L_GAN}
\end{equation}
\noindent where $\sigma(\cdot)$ is the sigmoid and expectations are over minibatch samples. 

During training, the discriminator parameters $\psi$ are updated by minimizing $L_D$, which encourages it to assign higher relative scores to real samples and lower scores to generated ones. 
The surrogate operator was trained using a combination of $L^1$ loss, perceptual loss, and adversarial loss ($L_G$) (\autoref{eq:L_GAN}). 
$L^1$ loss simply penalizes the difference between the $u_{\mathrm{tar}}$ and $u_f$, and guides the learning of the low-wavenumber structures.  
Perceptual loss is computed on intermediate feature maps from a pre-trained network.
Let $\{\phi_l\}_{l=1}^{L}$ denote the first $L$ convolutional layers of a pre-trained network. 
The perceptual loss is defined as $\sum_{l=1}^{L} w_{l} \; \big\| \phi_{l}(u_{\mathrm{tar}}) - \phi_{l}(u_f) \big\|_{1}$.
For 2D data we use the first five layers of VGG-19 \cite{simonyan2014very} with weights $(w_1,\ldots,w_5)=(0.1,\,0.1,\,1,\,1,\,1)$. 
For 3D data we use the first three layers of 3D-ResNet10 from Med3D \cite{chen2019med3d} with weights $(w_1,\,w_2,\,w_3)=(0.1,\,0.1,\,1)$.
Adversarial loss $L_G$ drives the time-conditioned neural operator to produce outputs that the discriminator considers more realistic than real samples on average.
The surrogate operator's parameters $\theta$ are updated by minimizing the composite loss term,
\begin{equation}
    \mathcal{L}_{\mathrm{advNO}} = \lVert u_{\mathrm{tar}} - u_f \rVert_{1} + \sum_{l=1}^{L} w_{l} \; \big\| \phi_{l}(u_{\mathrm{tar}}) - \phi_{l}(u_f) \big\|_{1} + \beta L_{G},
    \label{eq:L_advNO}
\end{equation} 
where \(\phi_l\) is the \(l^{th}\) feature map from a pretrained network, weighed by \(w_l\). $\beta$ is a hyper-parameter coefficient that controls the contribution of $L_G$ \cite{wang2021real} and was set to 0.1. 

The first term in \autoref{eq:L_advNO} guides the large-scale accuracy. 
Adversarial loss $L_G$ (third term), and the perceptual loss (second term) helps approximate the high-frequency turbulent structures without sacrificing the temporal coherence from the operator formulation. 
We alternate NO $(\theta \leftarrow \arg\min \mathcal{L}_{\text{advNO}})$ and discriminator $(\psi \leftarrow \arg\min L_{D})$ updates, progressively refining the NO by adversarially training it against an improving discriminator, during the course of training.

We also investigate a complementary strategy where we use generative models (VAE, DM, GAN) to super-resolve the smooth states estimated by the NO. 
The NO handles temporal evolution, while the generator restores fine-scale structure.
Next, we overview the different generative models used in this work.

\subsection*{Variational Autoencoders (VAE)}
We use an improved Vector-Quantized Variational Autoencoder (VQ-VAE) \cite{van2017neural}, adapted for super-resolution tasks. 
The unique aspect of VQ-VAE is that the continuous Gaussian latent space of a standard VAE is replaced with a discrete codebook of learnable embeddings. 
The finite set of embeddings serves as a learnable dictionary of features, which helps the model to compress similar inputs into the same code, improving generalization and interpretability. 
Moreover, the mapping to integer code yields a compact representation which also helps with memory constraints. The training of VQ-VAE is governed by the loss $\mathcal{L} $ defined as,
\begin{equation}
    \mathcal{L}_{VAE} = ||u_{\mathrm{tar}}-\mathrm{VAE}(\mathcal{G}_{\theta}(u_{\mathrm{in}}))||_{2}^{2} + ||\gamma[z_{e}(u_{\mathrm{tar}})]-e_k||_{2}^{2} + \beta||z_{e}(u_{\mathrm{tar}})-\gamma[e_k]||_{2}^{2},
\end{equation}
where $\mathrm{VAE}(\mathcal{G}_{\theta}(u_{\mathrm{in}}))$ is the state super-resolved by the VQ-VAE that takes NO predictions as its input. 
$z_{e}(u_{\mathrm{tar}})$ is the continuous latent representation mapped by the encoder, $e_{k}$ is the nearest codebook vector quantized from its continuous representation $z_{e}(u_{\mathrm{tar}})$, $\gamma[\cdot]$ is stop gradient function and $\beta$ is a hyperparameter balancing the encoder commitment. 
The stop-gradient function $\gamma[\cdot]$, prevents gradients from flowing through certain parts of the model, effectively treating them as constants during training. 
This is crucial in VQ-VAE because the quantized output is non-differentiable, as it's constructed using nearest-neighbor lookup. 
In the loss term $ ||\gamma[z_{e}(u_{\mathrm{tar}})]-e_k||_{2}^{2}$, the stop gradient function ensures that the encoder output $z_{e}(x)$ is kept constant, so that only the codebook vectors $e_k$ are updated.
In contrast, in the commitment loss term $\beta||z_{e}(u_{\mathrm{tar}})-\gamma[e_k]||_{2}^{2}$, the code book vectors are treated as constant, ensuring only encoder parameters are updated.
We observed that VQ-VAEs in their original form struggle to capture the fine-scaled turbulent structures. 
This occurs due to the compression and quantization during the codebook construction. 
Small oscillations become indistinguishable during nearest-neighbor lookup and hence, the decoder struggles to reconstruct high-wavenumber features solely from these quantized coarse latents. 
To address these issues, we use UNet-style skip connections to propagate the intermediate features from the encoder through concatenation at the decoder. 
This facilitates reconstructions of fine edges and small oscillations, relaxing the codebook quantized space bottleneck.
Importantly, this modification aligns well with our super-resolution objective, since the goal is not dimensionality reduction but accurate recovery of fine-scale turbulent features.

\subsection*{Diffusion Models (DM)}

We also consider a conditional score-based diffusion model to refine fields via denoising score matching across multiple noise levels. 
Let $u$ denote a field and $u_{\mathrm{cond}}$ the conditioning signal. 
The score network $s_{\phi}(u;\sigma,u_{\mathrm{cond}})$ estimates the gradient of the log-density of the Gaussian-perturbed conditional distribution,
\[
s_{\phi}(u;\sigma,u_{\mathrm{cond}})\;\approx\;\nabla_{u}\log p_{\sigma}\!\left(u \mid u_{\mathrm{cond}}\right),
\]
where $p_{\sigma}$ is the data distribution corrupted with $\mathcal{N}(0,\sigma^{2}I)$.
Following EDM\cite{karras2022elucidating}, we sample $\sigma$ from a log-normal schedule and minimize a preconditioned denoising score-matching objective over pairs $(u_{\mathrm{cond}},u_{\mathrm{tar}})$:
\[
\min_{\phi}\;\mathbb{E}_{\sigma,\,\varepsilon,\,(u_{\mathrm{cond}},u_{\mathrm{tar}})}
\left[
\omega(\sigma)\,\big\|\, s_{\phi}(u_{\mathrm{tar}}+\sigma\varepsilon;\sigma,u_{\mathrm{cond}}) + \tfrac{1}{\sigma}\varepsilon \big\|_{2}^{2}
\right],
\]
where $\varepsilon\!\sim\!\mathcal{N}(0,I)$ and $\omega(\sigma)$ is the EDM preconditioning weight. 
In other words, the score network sees the target corrupted by Gaussian noise at level $\sigma$ (i.e., $u_{\mathrm{tar}}+\sigma\varepsilon$) and is trained to predict the corresponding noise-scaled direction $-\varepsilon/\sigma$ that removes this corruption.
For super-resolution, the conditioning signal $u_{\mathrm{cond}}=\mathcal{G}_{\theta}(u_{\mathrm{in}})$, the oversmoothed NO output and for flow reconstruction, $u_{\mathrm{cond}}$ is the sparse observation. 
We start sampling from a Gaussian noise at a large $\sigma_{0}$ and progressively lower $\sigma\!\to\!0$, updating the sample with the learned score at each step.
We use EDM’s noise schedule and a Heun-style sampler for efficiency and stability. 
Training across a log-spaced range of $\sigma$ yields multi-scale fidelity, where large $\sigma$ captures coarse structure and small $\sigma$ restores high-wavenumber detail, but at the cost of iterative sampling.

\subsection*{Generative Adversarial Network (GAN)}
The GAN implemented in our work is based on Real-ESRGAN \cite{wang2021real}, which is an extension to the ESR-GAN \cite{wang2018esrgan}, designed to achieve enhanced super-resolution performance compared to previously developed GAN-based models. 
A deep convolutional upsampling network with residual-in-residual dense blocks (RRDB) \cite{wang2018esrgan} is used as the generator, and a UNet architecture is used as the discriminator. 
The RRDB stacks three Residual Dense Blocks, each with five $3\times3$ convolutions.
These nested convolutions substantially increase compute, explaining the higher FLOPs we report for NO+GAN in \autoref{tab:cost_accuracy_sr}.
For training, we follow the relativistic average GAN (RaGAN) objective described earlier, and include a VGG-19-based perceptual loss. 
RaGAN compares the relative realness of real vs. generated samples, which sharpens textures and stabilizes adversarial updates.
A UNet discriminator with spectral normalization is used to increase discriminator capacity and stabilize training.
Perceptual loss is computed on VGG-19 features before activation, together with an $L^1$ content loss to learn large-scale structures.

\subsection*{Extending from 2D to 3D: Memory-efficient network adaptations}

For the spatio-temporal super-resolution problem, the perceptual loss in adv-NO and GAN was computed on the 2D feature representations projected by the pretrained layers of VGG-19, which is a 2D convolutional network as discussed earlier. 
For forecasting three-dimensional isotropic turbulence with adv-NO and reconstructing the three-dimensional cylinder wake with GAN, we utilize a pre-trained 3D ResNet from the Med3D \cite{chen2019med3d}, which was originally trained on biomedical images of human organs. 
Specifically, we used the first three pre-trained layers of 3D-ResNet10 from \cite{chen2019med3d} with weights of $\{0.1,0.1,1\}$. 
However, we encountered out-of-memory errors.
To make our implementation memory-efficient, we modified how we utilized the discriminator. 
Instead of sending the full $128^3$ or $256\times256\times32$ dimensional snapshots for HIT and cylinder cases, respectively, as the input to the UNet-based discriminator and pretrained 3D-ResNet10, we utilized randomly sampled smaller sub-domains of size $32^3$ to compute the adversarial and perceptual losses. 
This strategy helped us avoid out-of-memory errors.
Our 3D implementation of the diffusion model\cite{karras2022elucidating} for the cylinder wake reconstruction also encountered out-of-memory errors.
In this case, the culprit was the linear and full attention blocks in the neural network architecture, which we turned off to make the network more memory efficient.

\subsection*{Training and implementation details}

{
All models used in the spatio-temporal super-resolution of 2D Schlieren images contained approximately two to four million trainable parameters, while those used for 3D forecasting and sparse flow reconstruction contained approximately six million parameters.
We intentionally selected relatively small models to ensure efficient training on a single GPU and to enable effective learning in the low-data regime.

All models were trained using the Adam optimizer \cite{kingma2014adam} with a learning rate of $10^{-4}$, and training was performed until convergence, with a maximum wall-clock time of 48 hours per experiment.
For the conventional NO trained solely to minimize the $L^2$ loss, we additionally employed a linear warm-up followed by a cosine annealing learning rate scheduler \cite{gupta2022towards} to accelerate convergence.
While this setup enabled faster training, the conventional NO consistently suffered from spectral bias, as discussed earlier.
To mitigate this limitation, we further trained the NO adversarially (adv-NO) and also evaluated VAE-, GAN-, and diffusion-based frameworks for super-resolving the NO-predicted states.
In general, adv-NO, VAE, GAN, and diffusion models exhibited slower convergence compared to conventional NOs but achieved superior reconstruction of fine-scale structures and turbulent dynamics as reported in \autoref{fig:genai_comparison}.
All training and inference were conducted on a single NVIDIA H100 GPU using the PyTorch framework \cite{paszke2019pytorch}.
}

\newpage
\section*{Acknowledgements}
The authors acknowledge Dr. Zhen Zhang, Prof. Martin Maxey and Prof. Johannes Brandstetter for the insightful discussions.

\noindent \textbf{Funding:} 
We acknowledge the funding from MURI/AFOSR FA9550-20-1-0358 project, the DOE-MMICS SEA-CROGS DE-SC0023191 award and DARPA-ABAQuS program grant number HR00112490526.
We also acknowledge the computational resources and services at the Center for Computation and Visualization (CCV), Brown University.

\noindent \textbf{Author Contributions:} 
V.O.: 
Conceptualization, Methodology, Software, Formal Analysis, Investigation, Data Curation, Writing - Original Draft, Writing - Review \& Editing and Visualization.
S.K.:
Methodology, Software, Formal Analysis, Investigation, Writing - Original Draft and Writing - Review \& Editing
A.B.:
Methodology, Software, Formal Analysis, Investigation, Writing - Original Draft and Writing - Review \& Editing
Z.W.:
Software, Investigation, Validation, Resources, Data Curation, Writing - Original Draft and Writing - Review \& Editing
G.E.K.:
Conceptualization, Validation, Formal Analysis, Resources, Writing - Review \& Editing, Supervision, Project administration and Funding acquisition

\noindent \textbf{Competing Interests:} 
The authors declare no competing interests. 

\noindent \textbf{Data and Code Availability:} 
The data and code will be made available after acceptance at our \href{https://github.com/vivekoommen/Gen4Turbulence}{GitHub Repository}\cite{Gen4Turbulence2025}

\bibliographystyle{unsrt}  
\bibliography{references}  
\newpage

\appendix
\counterwithin{figure}{section}
\setcounter{figure}{0}

\section{Analysis of spectral bias in Neural Operators}
\label{sec:sb_analysis}

We perform a frequency-resolved analysis of training dynamics in the operator-learning regime, assuming physically realistic target spectra and initializations whose Fourier sensitivity decays with frequency. 
Our analysis is architecture-agnostic and shows (i) global concentration of gradient mass on low wavenumbers at initialization and (ii) an explicit high-frequency tail-energy deficit over realistic training horizons.
\newline

\noindent\textbf{Lemma:}
Let $\Omega\subseteq\mathbb{R}^d$ be bounded and $f\in L^2(\Omega)$. We extend $f$ by zero to $\mathbb{R}^d$ while keeping the same notation $f$, and we use the unitary Fourier transform given by
\begin{equation}
    \widehat{g}(k)=(2\pi)^{-d/2}\int_{\mathbb{R}^d} g(x)\,e^{-ik\cdot x}\,dx,
\qquad \|g\|_{L^2(\mathbb{R}^d)}=\|\widehat{g}\|_{L^2(\mathbb{R}^d)}.
\end{equation}
For $\theta\in\mathbb{R}^p$ let $f_\theta$ denote the model output and $\mathcal{L}(\theta)$ be the associated $L^2$ loss given by
\begin{equation}
    \mathcal{L}(\theta)\;=\;\|f-f_\theta\|_{L^2(\Omega)}^2
\;=\;\int_{\mathbb{R}^d}\big|\widehat{f}(k)-\widehat{f_\theta}(k)\big|^2\,dk.
\end{equation}
Let $S_\theta(k)=\big\|\nabla_\theta\,\widehat{f_\theta}(k)\big\|$. We assume that at initialization $\theta = \theta_0$ and 
\begin{enumerate}
\item[(a)] $\widehat{f_{\theta_0}}\in L^2(\mathbb{R}^d)$ (e.g.\ $\widehat{f_{\theta_0}}\equiv 0$), and
\item[(b)] $S_{\theta_0}\in L^2(\mathbb{R}^d)$ with vanishing tail:
$\displaystyle \int_{|k|>K} S_{\theta_0}(k)^2\,dk \to 0$ as $K\to\infty$.
\end{enumerate}
Then
\begin{enumerate}
\item[(i)] For every $\varepsilon>0$ $\exists$ $K_\varepsilon<\infty$ such that
\begin{equation}
    \big\|(\nabla_\theta \mathcal{L}(\theta_0))_{\le K_\varepsilon}\big\|
\;\ge\; (1-\varepsilon)\,\big\|\nabla_\theta \mathcal{L}(\theta_0)\big\|,
\end{equation}
where $(\nabla_\theta \mathcal{L}(\theta))_{\le K}$ denotes the contribution to
\begin{equation}
    \nabla_\theta \mathcal{L}(\theta)=2\,\Re\!\int_{\mathbb{R}^d}\big(\widehat{f_\theta}(k)-\widehat{f}(k)\big)\,
\overline{\nabla_\theta \widehat{f_\theta}(k)}\,dk
\end{equation}
for $\{|k|\le K\}$.

\item[(ii)] For every $\theta$ and $K\ge 0$,
\begin{equation}
    \int_{|k|>K}\big|\widehat{f_\theta}(k)\big|^2\,dk
\;\le\; 2\,\mathcal{L}(\theta)\;+\;2\int_{|k|>K}\big|\widehat{f}(k)\big|^2\,dk.
\end{equation}
\end{enumerate}
\noindent \textbf{Sketch of the Proof:} \newline

\noindent From unitary fourier transfrom we have
\begin{equation}
    \widehat{g}(k)=(2\pi)^{-d/2}\int_{\mathbb{R}^{d}} g(x)\,e^{-ik\cdot x}\,dx,
\qquad \|g\|_{L^2(\mathbb{R}^{d})}=\|\widehat{g}\|_{L^2(\mathbb{R}^{d})}.
\end{equation}Now using Parseval-Plancherel identity \cite{parseval1806memoire}, we can write Eqn (2) as
\begin{align}
    \mathcal{L}(\theta) &= \int_{\mathbb{R}^d}\big|\widehat{f}(k)-\widehat{f_{\theta}}(k)\big|^{2}\,dk \\
&= \int_{\mathbb{R}^d}\big(\widehat{f}(k)-\widehat{f_{\theta}}(k)\big)\,\overline{\big(\widehat{f}(k)-\widehat{f_{\theta}}(k)\big)}\,dk.
\end{align}Let $f_\theta$ be differentiable (Frechet/Gateaux) in $\theta$, then based on assumption (b), for any direction $\nu \in \mathbb{R}^p$, by chain rule we have
\begin{equation}
    D_\nu \mathcal{L}(\theta) = 2\,\Re\!\left\langle f_\theta - f,\, D_\nu f_\theta \right\rangle_{L^2(\mathbb{R}^d)} = 2\,\Re\!\left\langle \widehat{f_\theta} - \widehat{f},\, D_\nu \widehat{f_\theta} \right\rangle_{L^2(\mathbb{R}^d)}.
\end{equation}Now using Riesz representation, Eqn (9) is equivalent to 
\begin{equation}
    \nabla_{\theta} \mathcal{L}(\theta) = 2 \int_{\mathbb{R}^{d}} \Re \left[(\hat{f}_{\theta}(k)-\hat{f}(k))\overline{\nabla_{\theta}\hat{f}_{\theta}(k)}\right]dk.
\end{equation}Now let, for $\theta=\theta_{0}$ and for a cutoff $K\geq0$,  we have
\begin{align}
    g_{\leq K}(\theta_{0}) & = 2\Re \int_{|k|\leq K}(\hat{f_{\theta_{0}}}(k)-\hat{f}(k))\overline{\nabla_{\theta}\hat{f_{\theta_{0}}}(k)}dk, \\
    g_{> K}(\theta_{0}) & = 2\Re \int_{|k|> K}(\hat{f_{\theta_{0}}}(k)-\hat{f}(k))\overline{\nabla_{\theta}\hat{f_{\theta_{0}}}(k)}dk.
\end{align} From Eqn. (10) we have 
\begin{equation}
    \nabla_{\theta}\mathcal{L}(\theta_{0}) = g_{\leq K}(\theta_{0})+g_{>K}(\theta_{0}).
\end{equation}Therefore,
\begin{equation}
    \big\|g_{\leq K}(\theta_{0})\big\|\;\ge\; \big\|\nabla_{\theta}\mathcal{L}(\theta_{0})\big\|- \big\|g_{>K}(\theta_{0})\big\|.
\end{equation}Now using triangle inequality in $\mathbb{R}^{p}$ and $|\Re z|\leq |z|$,
\begin{equation}
\big\|g_{>K}(\theta_{0})\big\|
\le 2\int_{|k|>K}\big|\widehat{f_{\theta_{0}}}(k)-\widehat{f}(k)\big|\;
\big\|\nabla_{\theta}\widehat{f_{\theta_{0}}}(k)\big\|\,dk.
\end{equation}Now applying the Cauchy-Schwarz inequality
\begin{equation}
    ||g_{>K}(\theta_{0})||\leq 2||\hat{f_{\theta_{0}}}-\hat{f}||_{L^2 (|k|>K)}||S_{\theta_{0}}||_{L^2 (|k|>K)}.
\end{equation}
From (a) and (b) we have
\begin{equation}
\ \big\|\widehat{f_{\theta_{0}}}-\widehat{f}\big\|_{L^2(|k|>K)} \to 0
\quad\text{and}\quad
\big\|S_{\theta_{0}}\big\|_{L^2(|k|>K)} \to 0
\qquad (K\to\infty).
\end{equation} Hence,  $2||\hat{f_{\theta_{0}}}-\hat{f}||_{L^2 (|k|>K)}||S_{\theta_{0}}||_{L^2 (|k|>K)} \rightarrow 0$, and therefore
\begin{equation}
\exists\,K_{\varepsilon}<\infty\ \text{s.t.}\ 
\big\|g_{>K_{\varepsilon}}(\theta_{0})\big\|
\le \varepsilon\,\big\|\nabla_{\theta}\mathcal{L}(\theta_{0})\big\|.
\end{equation}From Eqn. (14) and (16) we have 
\begin{align}
\big\|g_{\le K_{\varepsilon}}(\theta_{0})\big\|
= \big\|(\nabla_{\theta}\mathcal{L}(\theta_{0}))_{\le K_{\varepsilon}}\big\|
\;\ge\; (1-\varepsilon)\,\big\|\nabla_{\theta}\mathcal{L}(\theta_{0})\big\|.
\end{align}We have,
\begin{align}
    |a|-|b|\le |a-b|, \qquad
   |a|^{2}\le 2\,|a-b|^{2}+2\,|b|^{2}.
\end{align} Now using  Eqn. (23) and a = $\hat{f_{\theta}(k)}$ and b = $\hat{f}(k)$ and then integrating over $|k| > K$ we have,
\begin{equation}
\int_{|k|>K}\big|\widehat{f_{\theta}}(k)\big|^{2}dk 
\le 2\int_{|k|>K}\big|\widehat{f_{\theta}}(k)-\widehat{f}(k)\big|^{2}dk 
+ 2\int_{|k|>K}\big|\widehat{f}(k)\big|^{2}dk.
\end{equation}
Now from Eqn. (8) and (25), we have
\begin{equation}
\int_{|k|>K} \big|\widehat{f_{\theta}}(k)\big|^{2}\,dk
\;\le\;
2\int_{\mathbb{R}^{d}} \big|\widehat{f_{\theta}}(k)-\widehat{f}(k)\big|^{2}\,dk
\;+\; 2\int_{|k|>K} \big|\widehat{f}(k)\big|^{2}\,dk
\;=\; 2\,\mathcal{L}(\theta) + 2\int_{|k|>K} \big|\widehat{f}(k)\big|^{2}\,dk.
\end{equation}

\clearpage
\section{Super-resolving Schlieren images of an impinging jet - Additional Results}
\label{sec:imp_jet_bicubic}

{
\autoref{fig:impjet} compares the superresolved states estimated through bicubic interpolation, conventional NO trained to minimize the Euclidean error, and the adversarially trained NO (adv-NO). 
The bicubic interpolation and NO suffer from spectral bias/over-smoothing.
The energy spectrum plots in the last row clearly indicate that the extent of spectral bias is worse in bicubic interpolation compared to NO. 
Furthermore, at $t+3\tau$, where even the low resolution input is non-existent, the low-wavenumbers estimated through bicubic interpolation does not align with the reference.
Whereas, the NO is able to learn the dynamics of at least the low wavenumbers correctly. 
However, the Euclidean error-based training causes the NO to remain completely oblivious to the dynamics at the higher wavenumbers. 
This limitation can be avoided altogether through adversarial training of the NO, which is able to estimate states with almost perfectly aligned energy spectra.  
}

{
\begin{figure}[h!]
  \centering
  \includegraphics[width=0.8384\linewidth]{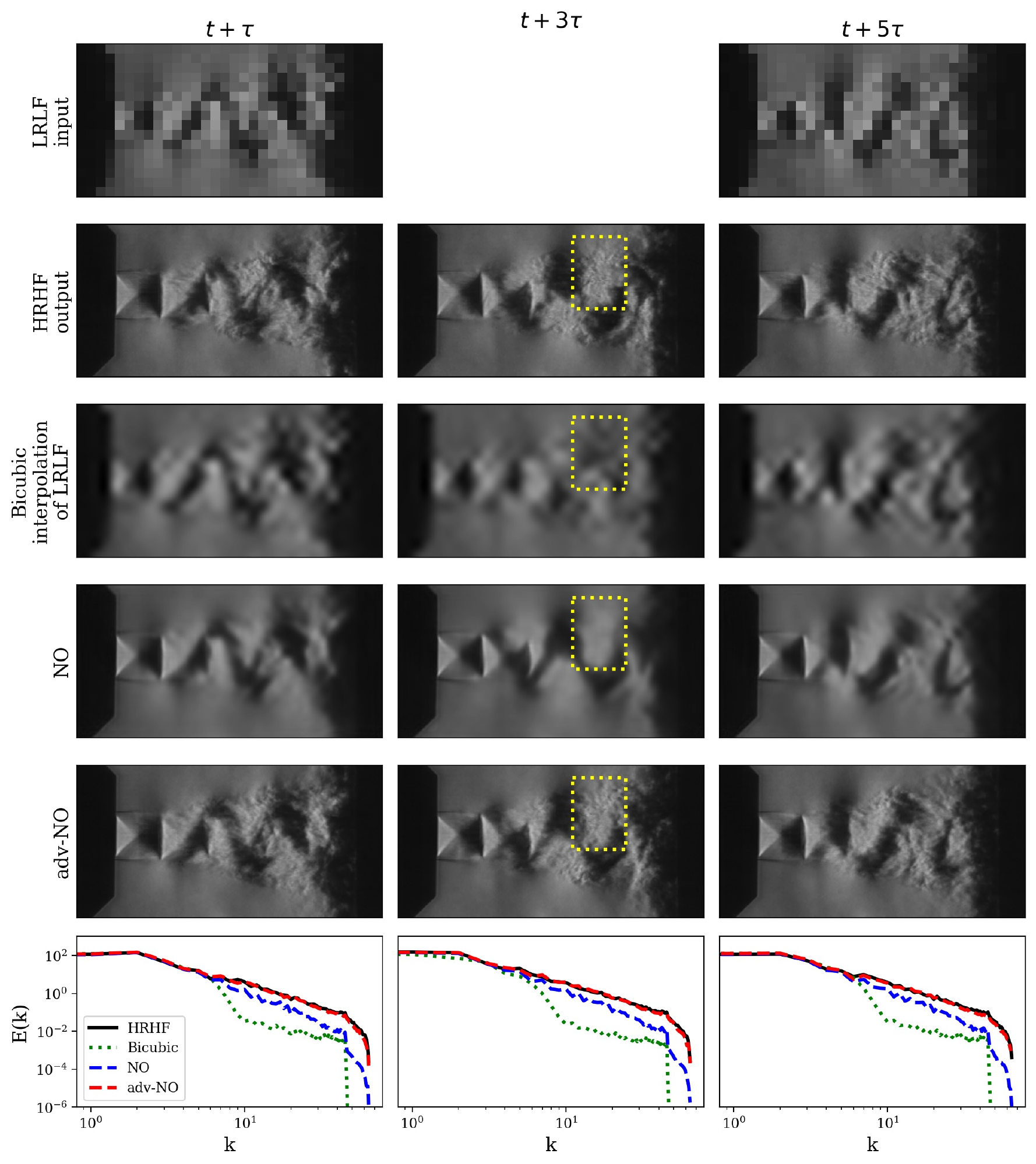}
  \caption{\textbf{Impinging jet.} Super-resolution of low-resolution, low-frame-rate (LRLF) Schlieren visualizations of an impinging jet. 
  The model reconstructs high-resolution, high-frame-rate (HRHF) density gradients, by upscaling the spatial and temporal resolutions by a factor of 8 and 4 respectively.
  The last row compares the energy spectrum of the density gradient reconstructed through bicubic interpolation, conventional neural operator (NO), and adversarially trained neural operator (adv-NO).
  }
  \label{fig:impjet}
\end{figure}
}

\clearpage
\section{Forecasting Forced Homogeneous Isotropic Turbulence - Additional Results}
\label{supple_sec:forecast}

{
\begin{figure}[h!]
  \centering
  \includegraphics[width=0.99\linewidth]{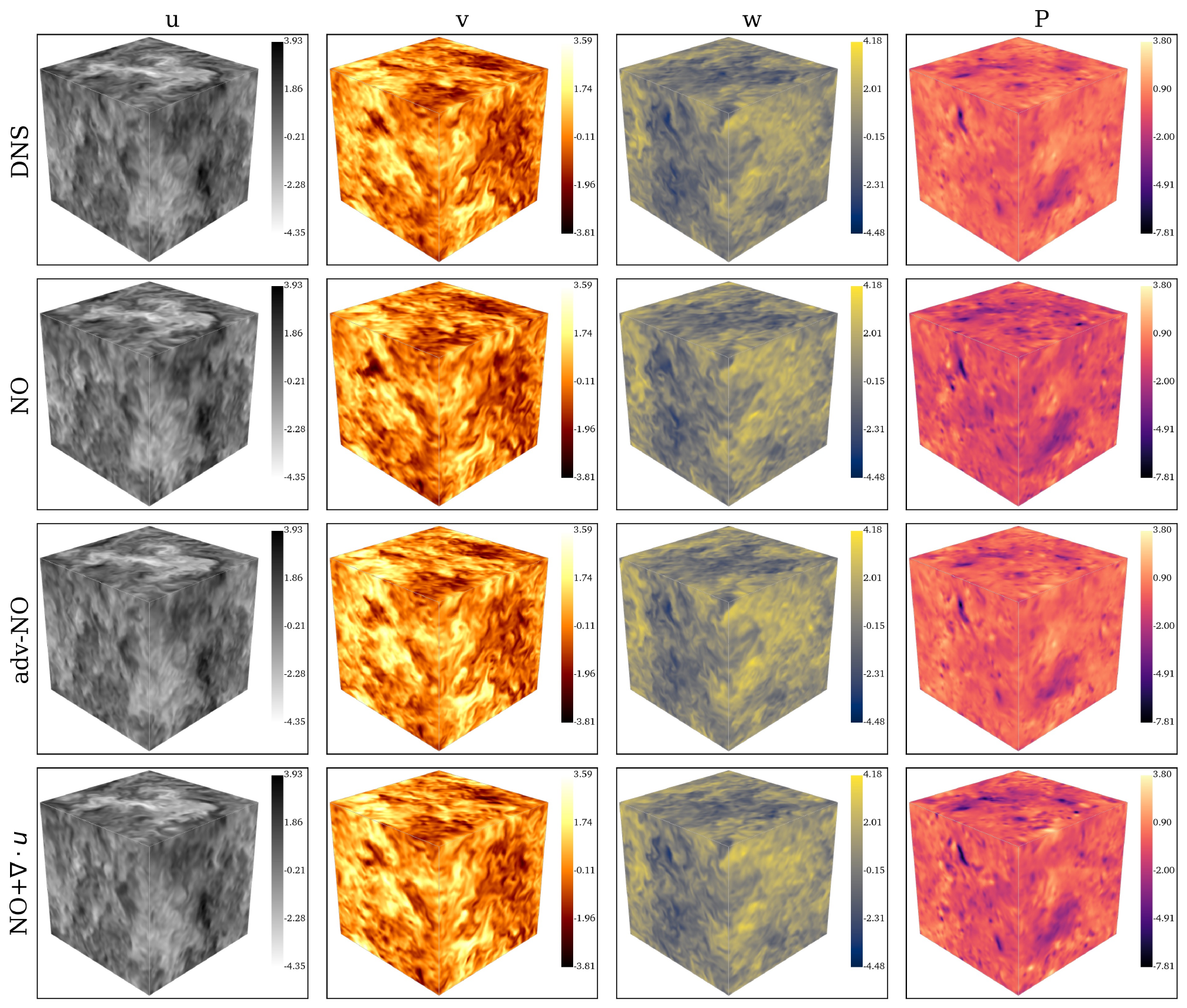}
  \caption{\textbf{Forecasting Homogeneous Isotropic Turbulence (HIT) ($Re_{\lambda}=90$).}
  The velocity and the pressure states simulated using spectral element method is used as a reference for comparing the predictions of NO, adv-NO and a physics-informed variant of the neural operator that penalizes the divergence of the velocity vector.
  All the models were trained only on 160 timesteps of a single trajectory.
  The states at $\Delta t=2t_E$ ($t_E$ is the eddy turnover time) is visualized. 
  }
  \label{fig:hit3d_uvwP}
\end{figure}
}

{
\begin{figure}[h!]
  \centering
  \includegraphics[width=0.99\linewidth]{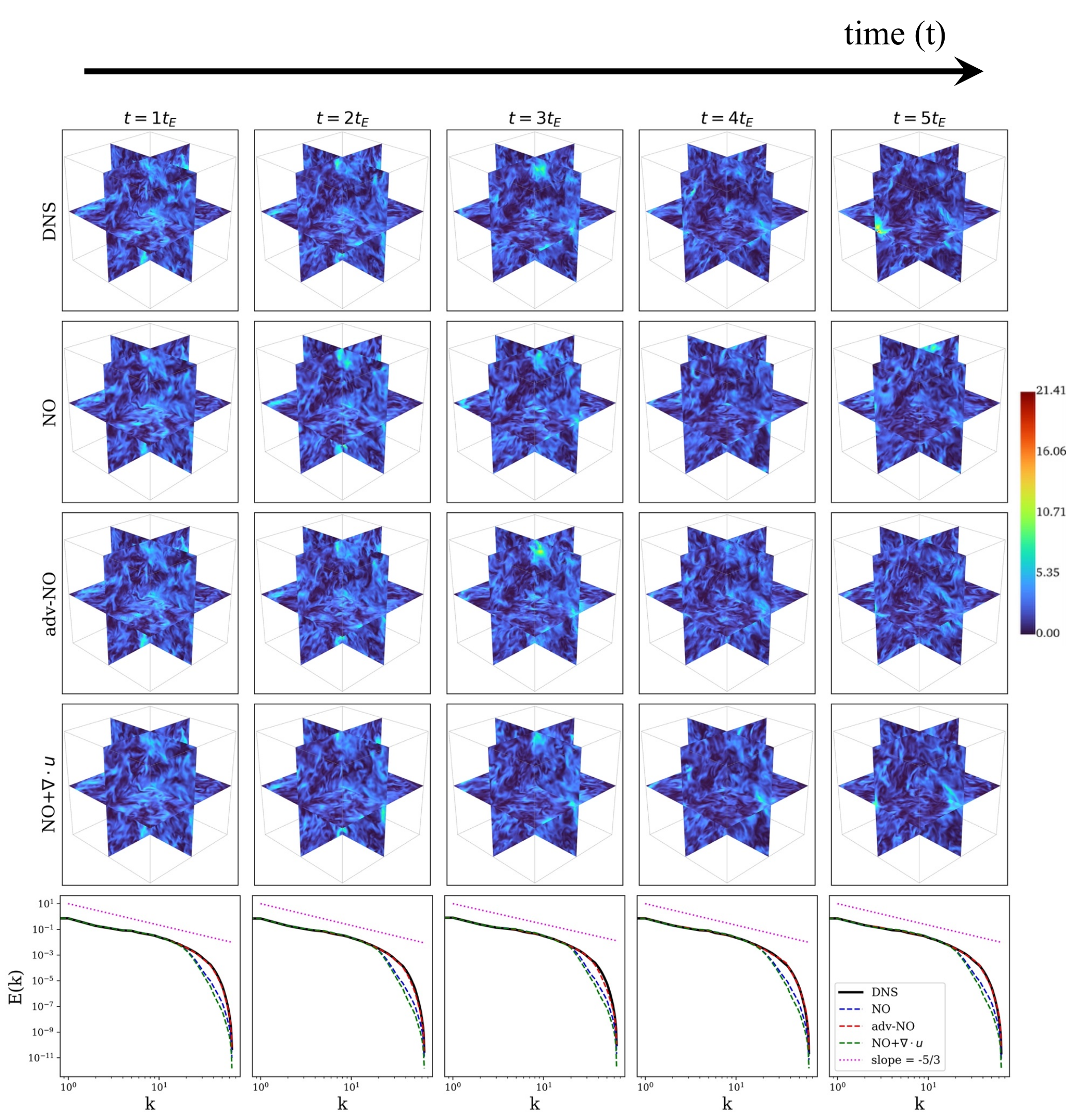}
  \caption{\textbf{Evolution of kinetic energy (KE) in HIT ($Re_{\lambda}=90$).}
  We show the time evolution of KE by visualizing the mid xy, yz, zx planes. 
  The columns represent the time evolution from one to five eddy turnover time-scales ($t_E$)
  The last row shows the energy spectrum as it evolves in time.
  }
  \label{fig:hit3d_ke_time}
\end{figure}
}

\clearpage
\section{Flow reconstruction from partial observations}

{
\begin{algorithm}[H]
\caption{Preparing Masked Inputs for Training}\label{alg:mask_prep_training}
\begin{algorithmic}[1]
\State \textbf{Input}: full flow tensor $x \in \mathbb{R}^{\text{bs} \times 4 \times n_x \times n_y \times n_z}$
\State Sample mask fraction $\alpha \sim \mathcal{U}(0,1)$
\If{rand() < 0.5}
  \State \textbf{Independent masking}:
  \For{each sample $b$ in mini-batch}
    \State Generate random mask $m_b \in \{0,1\}^{4 \times n_x \times n_y \times n_z}$ with fraction $\alpha$ zeros
  \EndFor
\Else
  \State \textbf{Shared mask across fields}:
  \For{each sample $b$ in mini-batch}
      \State Generate one mask $m \in \{0,1\}^{1 \times n_x \times n_y \times n_z}$ with fraction $\alpha$ zeros
      \State $m_b \gets \text{repeat}(m, 4)$ 
  \EndFor
\EndIf
\State Masked input: $x_\text{masked} \gets x \cdot m$
\State Conditional input: $x_\text{in} \gets \text{concat}(x_\text{masked}, m)$
\State \textbf{return} $x_\text{in}$
\end{algorithmic}
\end{algorithm}
}

{
\begin{figure}[h!]
  \centering
  \includegraphics[width=0.9\linewidth]{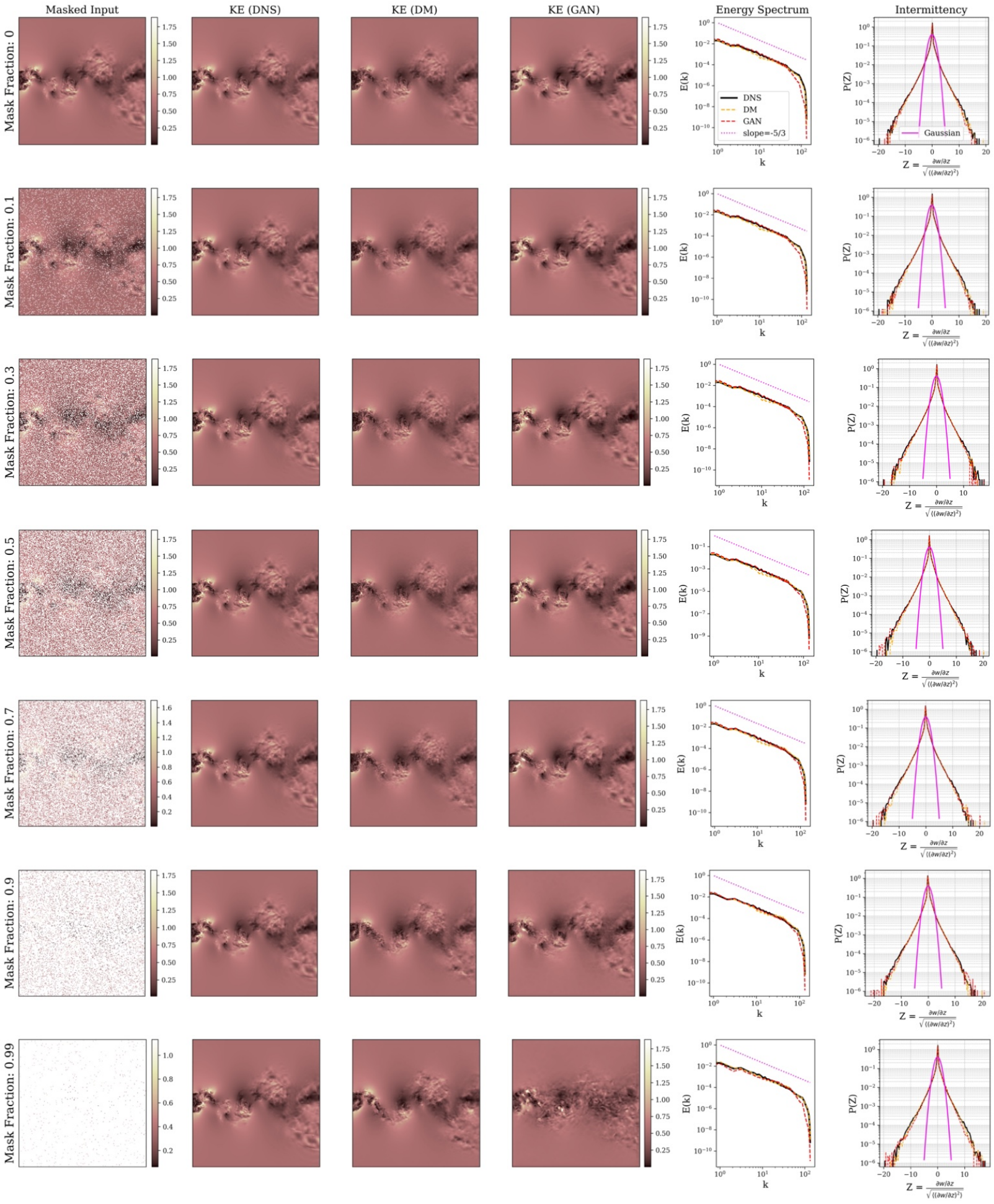}
  \caption{\textbf{Zero-shot flow reconstruction: Sensitivity to mask fraction.}
   The masked input, kinetic energy of the reference DNS, kinetic energy of the velocity fields reconstructed by Diffusion Model and GAN, their energy spectra and intermittency curves are visualized for different mask fractions.
   Both DM and GAN are able to successfully reconstruct the vortex street for all mask fractions (except GAN when mask fraction=0.99)
  }
  \label{fig:FR_TS1}
\end{figure}
}

{
\begin{figure}[h!]
  \centering
  \includegraphics[width=0.8\linewidth]{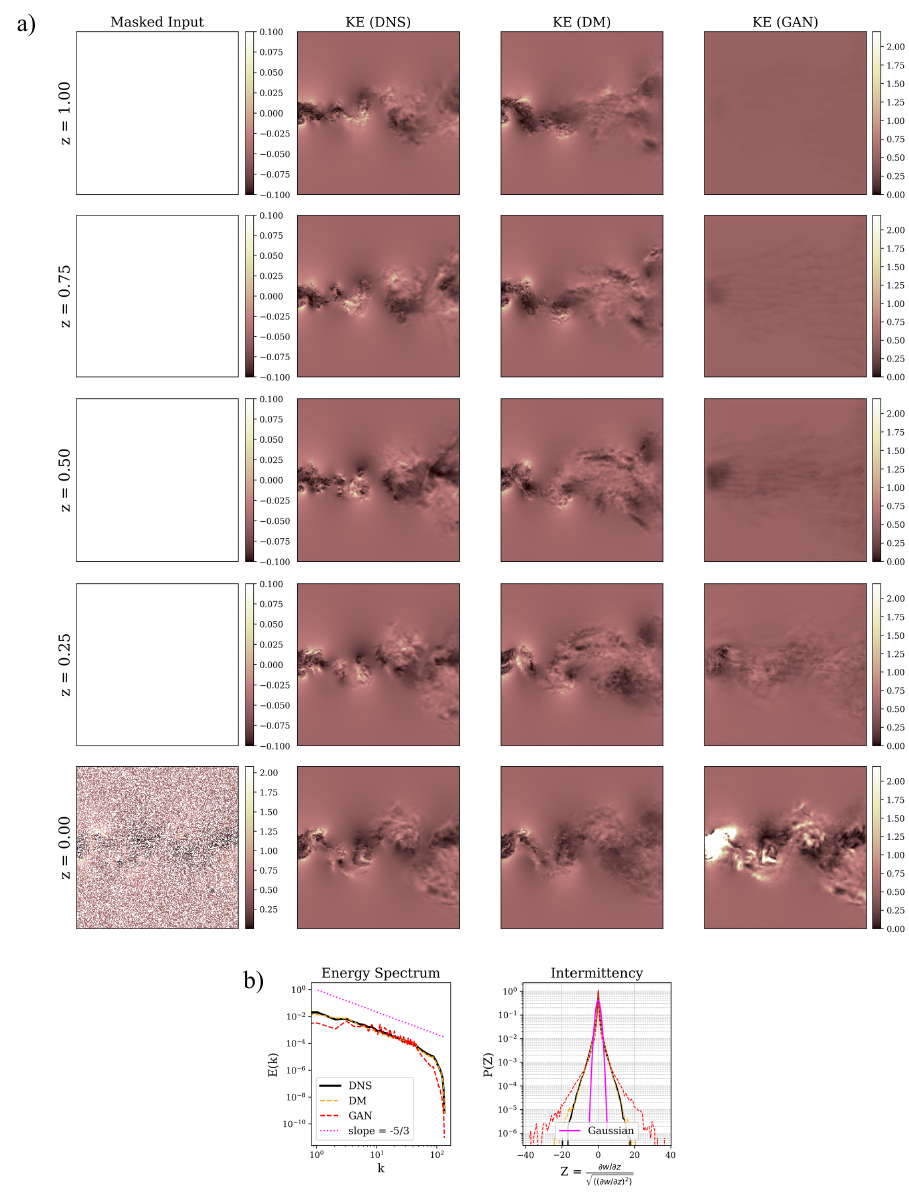}
  \caption{\textbf{Test setup 2 – reconstruction of the full 3D flow field from sparse velocity measurements on a single plane ($z=0$).}
  a) Kinetic energy (KE) from DNS and KE reconstructed by the diffusion model (DM) and GAN, at different planes.
  b) Comparison of the spatially averaged energy spectrum and velocity intermittency between DNS, DM, and GAN reconstructions.
  }
  \label{fig:FR_TS2}
\end{figure}
}

{
\begin{figure}[h!]
  \centering
  \includegraphics[width=0.8\linewidth]{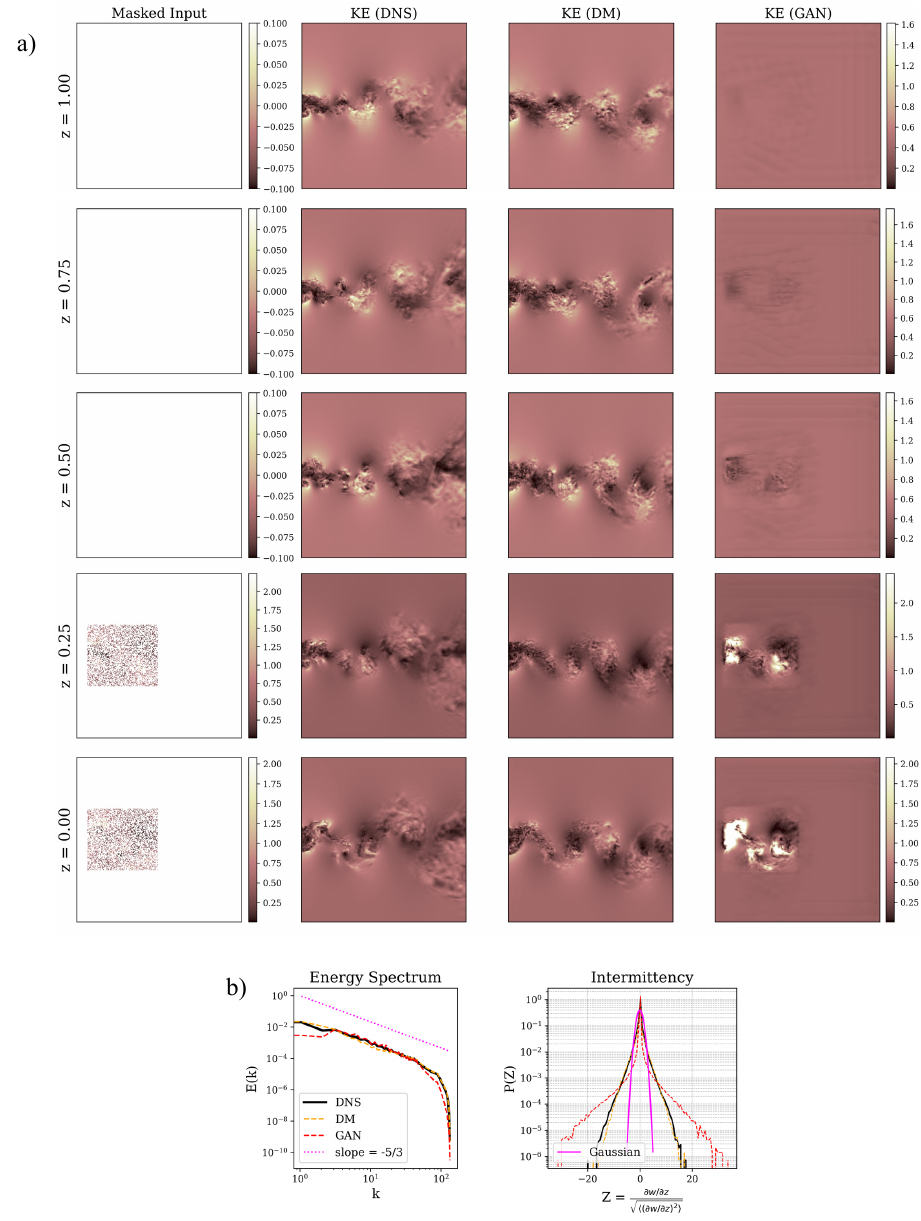}
  \caption{\textbf{Test setup 3 – reconstruction of the full 3D flow field from sparse velocity measurements in a sub-domain.}
  a) Kinetic energy (KE) from DNS and KE  reconstructed by the diffusion model (DM) and GAN, at different planes.
  b) Comparison of the energy spectrum and velocity intermittency between DNS, DM, and GAN reconstructions.
  }
  \label{fig:FR_TS3}
\end{figure}
}

{
\begin{figure}[h!]
  \centering
  \includegraphics[width=0.8\linewidth]{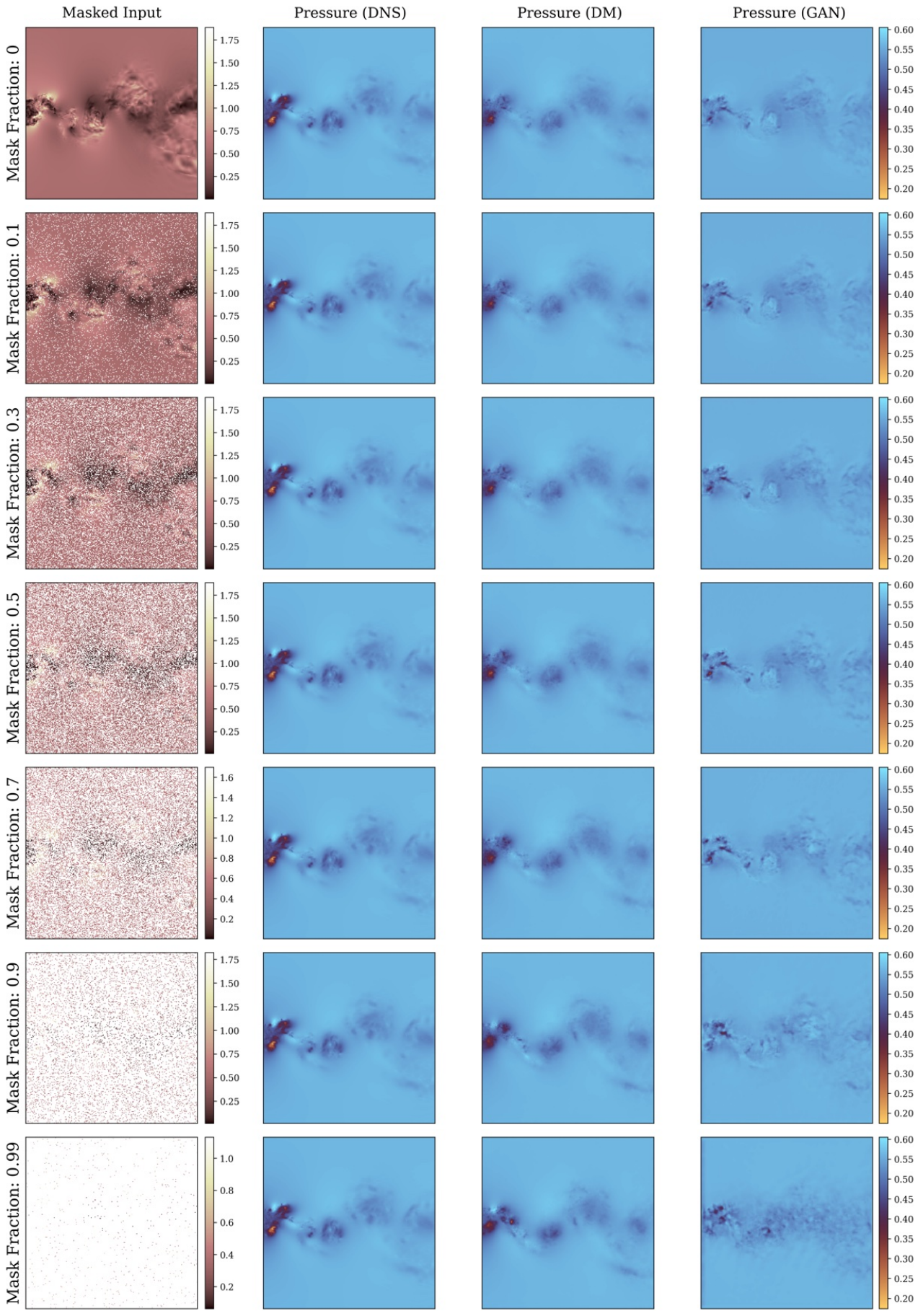}
  \caption{\textbf{Test setup 4 – reconstruction of the full 3D pressure field from sparse velocity measurements.}
  Pressure field from DNS and those reconstructed by the diffusion model (DM) and GAN, at the mid-plane slice, for different mask fractions.
  }
  \label{fig:FR_TS4}
\end{figure}
}

\end{document}